\documentclass{elsart}
\usepackage{amssymb}
\usepackage{graphicx}

\newlength\figwidth
\newlength\spherewidth
\newcommand{\TwoSides}[2]
{\begin{minipage}{\spherewidth}
\centering
\includegraphics*[angle=270,width=\textwidth]{#1.eps}
\end{minipage}\hfill
\begin{minipage}{\spherewidth}
\centering
\includegraphics*[angle=270,width=\textwidth]{#2.eps}
\end{minipage}}

\begin{document}
\begin{frontmatter}
\title{A Diffusion Model for Classical Chaotic Compound Scattering}
\author[cic,fisica]{F. Leyvraz},
\author[cic,grenoble]{M. Lombardi\thanksref{Corresponding}}
 \and
\author[cic,fisica]{T.H. Seligman}
\address[cic]{Centro Internacional de Ciencias, Cuernavaca, Mexico}
\address[fisica]{Centro de Ciencias F\'\i sicas, UNAM, Av. Universidad
s/n, Col. Chamilpa, Morelos, Mexico}
\address[grenoble]{Laboratoire de Spectrom\'etrie Physique
(CNRS UMR 5588),
Universit\'e Joseph-Fourier de Grenoble, BP87,
F-38402 Saint~Martin~d'H\`eres~C\'edex, France}
\thanks[Corresponding]{Corresponding author; Fax: +33~476~514~544;
e-mail: Maurice.Lombardi@ujf-grenoble.fr}
\begin{abstract}
 We consider the classical map proposed previously to be the exact
classical analog
of Rydberg Molecules calculated with the approximations relevant
to the multi-channel quantum defect theory.
The resulting classical map is analyzed at energies above the
threshold for the Rydberg electron.
At energies very near to this threshold we find the possibility
of bounded motion for positive energy due to conserved tori as well
as the possibility of forming a compound system, i.e. a system
where the particle is trapped for long times before emerging
again to the continuum. The compound scattering displays unusual
features for short time behavior. A diffusion model explains these
features.
\end{abstract}
\end{frontmatter}

\section{Introduction}

The Multichannel Quantum Defect Theory (MQDT) is a well
established quantum approximation framework
to accurate spectroscopic studies of atoms \cite{Seaton}
and molecules \cite{Fano}
near ({\it i.e.} both below and above) an ionization limit.
More recently the MQDT theory for molecules has been used as an
interesting tool to test ideas
about the quantum signatures of classical chaos
\cite{LombLabast,LombSelig,ComAtMol}.
Since this correspondence is valid only near the semi classical limit,
{\it i.e.} for large quantum numbers, some of these these quantum
numbers have
been taken much larger than their usual experimental values.
To simplify the mathematical analysis, an approximation
(the conservation of the angular momentum $L$ of the outer (Rydberg)
electron)
valid in the case of usual small molecules only for low $L$ has been kept
for much higher values of $L$. Indeed we apply a low-$L$ -- approximation 
in the semi-classical limit, i.e. for large $L$.
Other approximations better suited to actual
small molecules have been used in recent calculations \cite{CasShep}.
A possible experimental way out is to study much larger molecules,
and this is not out of reach since recent experimental
studies have found Rydberg states on molecules as large as
Benzene Argon complexes\cite{RydClust}, Diazacyclooctane (DABCO)
and bis (benzene) chromium (BBC)\cite{EvenLevine}.
These experimental studies however involve other issues,
namely the vibrational transfers between the very dense set of
vibrational
levels of such large molecules.
While all these experimental issues are very interesting to study,
we will not consider them in this paper.

We will use the model only for its interesting theoretical properties
with respect to the problem at hand, namely the study of chaotic scattering.
Indeed  one of us (M.L.) with co workers proposed
a classical map, the quantization of which
is exactly solved by the MQDT \cite{LombLabast}.
This approximate system deserves to be studied for its own sake
precisely because we can
compute the corresponding quantum results without approximation.
The study of the correspondence between classical and quantum results,
however, will be deferred to future work. In the present paper we establish
only interesting classical results.
With this proviso we shall in this paper refer to the approximate
classical system as the Rydberg molecule.

The Rydberg molecule provides a remarkably simple example of
a two dimensional system that displays compound behavior
and positive energy bound states typical of many-body systems.
It will therefore be of interest to study its scattering functions.

Among the classical scattering functions the time delays
are commonly used as they display the singularity structure
of such functions very clearly and simultaneously have
an intuitive meaning. The number of iterations of an internal
Poincar\'e map are a convenient alternative to time delays that does not
reflect distortions by far flung trajectories that are particularly
important in the Coulomb case. As we are precisely in  this situation
we shall mainly use the latter, though we shall also display the
mentioned distortions. The results for the number of iterations
display a very surprising structure of successive decay laws.
We cannot hope to obtain an understanding by
reconstructing the chaotic saddle.
Indeed we have not been able to resolve the fractal structure of
the scattering functions  and therefore no branching tree was obtained.
The reason was precisely the complexity induced by the far flung
trajectories, which produce much too narrow structures
(see the discussion at the beginning of 
sec.~\ref{sec:NumericalResults}).
We therefore develop a diffusive model that reflects this situation and
can be qualitatively related to the dynamical behavior of our
system.

\section{The Classical Model}

In ref.~\cite{LombLabast} the Rydberg molecule is described classically
as a
map of the unit sphere
onto itself. This sphere corresponds to the orientations of the
orbital angular momentum ${\vec L}$
of the electron (whose absolute value is approximated to be constant)
after each collision with the molecular core.
The construction of the map and its description turn out to be
somewhat involved. Conceptually, the simplest approach seems to be
that of looking at the system in the laboratory frame. 
The final results will be stated in a form that is essentially
frame independent.

\begin{figure}[hbp]
\begin{center}
\includegraphics[width=\figwidth,angle=270]{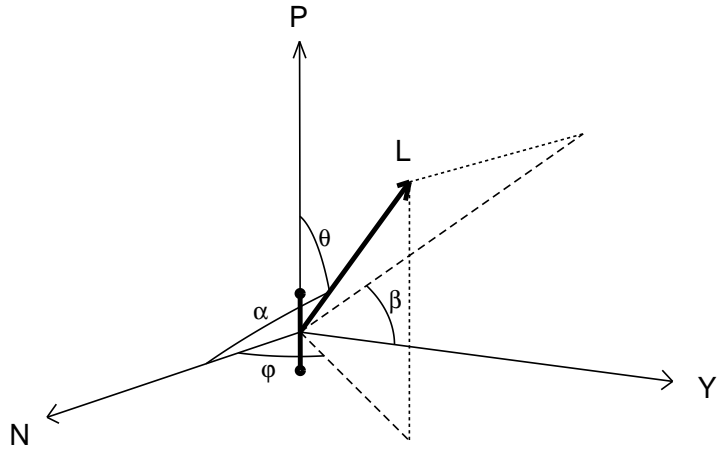}
\end{center}
\caption{Rydberg molecule reference frame.}
\label{fig:schema}
\end{figure}

We define a unit vector $\vec P$ in the direction of the core axis, 
the electron angular momentum $\vec L$, the core angular 
momentum $\vec N$ and the electronic energy $E_e$. This yields
a total of seven quantities. (We do not consider the quantities
$\vec P$, since they are determined up to an initial condition
by the time evolution of $\vec N$ which is perpendicular to
$\vec P$.) Between these seven quantities, 
there exist five relations given by total energy ($E$) conservation, the
conservation of total angular momentum
\begin{equation}
\vec J=\vec L+\vec N
\label{eq:total-J}
\end{equation} 
and finally, the approximate conservation
of the magnitude of the electronic angular momentum $L$. 
From this follows that one only needs two numbers to 
specify fully the configuration of the system once total energy,
total angular momentum and $L$ are known. For these we choose
the two polar angles $\theta$ and $\phi$ determining the relative 
orientation of $\vec L$ with respect to $\vec P$ as polar axis and
$\vec N$ as reference axis for $\phi$ or, alternatively,
the angles $\alpha$ and $\beta$ determining the orientation of 
$\vec L$ with respect to $\vec N$ and $Y$ (see fig.~\ref{fig:schema}).
As a matter of notation, we define $R_{\vec a}(\psi)$ to be a rotation
of angle $\psi$ around the axis $\vec a$ in the positive direction. 
We are now ready to define the two basic components
of the map: All quantities with an index $n$ refer to a given time
step, whereas primed quantities are intermediate quantities for
computation only. We therefore aim to express such quantities as 
$\vec L_{n+1}$ in terms of $\vec L_n$.
\begin{itemize}
\item[(1)]
Free Keplerian motion: When the electron is far from the core,
it feels only the isotropic $1/r$ potential of the core and thus follows
a free Coulomb trajectory, so that its angular momentum $\vec L$ is fixed.
The core on the other hand, feels no torque from the electron and thus
performs a free rotation. 
This implies for $\vec L$, $\vec N$
and $\vec P$ the following evolution:
\begin{eqnarray}
\vec L^\prime&=&\vec L_n\nonumber\\
\vec N^\prime&=&\vec N_n\nonumber\\
\vec P^\prime&=&R_{\vec N_n}\left(2\pi\frac{T_e}{T_N}\right)\vec P_n
\label{eq:free:motion}\\
T_e&=&2\pi\left(-2 E_e\right)^{-3/2}\nonumber\\
T_N&=&2\pi\left(2BN_n\right)^{-1}\nonumber
\end{eqnarray}
Here $T_e$ and $T_N$ represent the Keplerian period of the
electron and the free rotation period of the core respectively,
and $2B$ is the reciprocal of the moment of inertia of the core,
according to usual molecular spectroscopy notation. 
Atomic units ($e=\hbar=m=1$) are used throughout and
the equations given simply follow from the elementary
mechanical properties of Kepler motion and of free rotation. 

\item[(ii)] After the free motion, one has a ``collision'' of the
electron with the core, when it is near enough to feel 
the anisotropic part of its potential.
Due to cylindrical symmetry of this potential, the projection
$\Lambda = \vec L\cdot\vec P^\prime$ of $\vec L$ on the molecular 
axis $\vec P^\prime$ is conserved. Again, as pointed out above,
we assume that the magnitude of $\vec L$ is conserved. 
This collision is thus described by a precession of ${\vec L}$ around
the molecular axis $\vec P^\prime$ by an angle $\delta\phi$. Due to invariance
of the core potential in any plane containing the molecular
axis $\vec P^\prime$, $\delta\phi$ must be odd in $\theta \to \pi - \theta$.
We take in the present paper the simplest choice
$\delta\phi = K \cos\theta=K\vec L\cdot\vec P^\prime/L$, 
which defines the coupling strength $K$. In equations, this leads to
\begin{eqnarray}
\vec P_{n+1}&=&\vec P^\prime\nonumber\\
\vec L_{n+1}&=&R_{\vec P^\prime}(\delta\phi)\vec L^\prime\\
\vec N_{n+1}&=&\vec J-\vec L_{n+1}\nonumber
\end{eqnarray}
It should be noted that the effects of the change in $\vec N$
in the last equation are rather profound. First, and most important,
there is an energetic consequence. This change
involves a change in the magnitude $N$ of $\vec N$ and hence
in the core kinetic energy $E_N=B N^2$. This in turn reflects itself
in a change in $E_e = E-E_N$, which in turn, will affect both $T_N$ and $T_e$
in the following free motion. The system is therefore quite far from 
being a simple kicked top, in which the motion of $\vec L$ 
would be decoupled from the rest of the dynamics.
Second there is a kinematical consequence: since $\vec L$ has rotated
by $\delta\phi$ around $\vec P^\prime$, and thus moved with respect to $\vec J$,
this entails a reorientation of $\vec N$ and, since the angle $\phi$ is
defined with $\vec N$ as reference axis (see fig.~\ref{fig:schema}),
this produces an extra $\delta\phi^\prime$, which was called frame recoil
in ref.~\cite{LombLabast}
\end{itemize}
In the Appendix B of this ref.~\cite{LombLabast} are given explicit
expressions for the map established in molecular frame, which, even if less
obvious to visualize, are easier to write down explicitly.

In any case what we shall be always presenting are the values of 
$(\theta_n,\phi_n)$ describing the orientation of $\vec L$
with respect to $\vec P, \vec N$, which are equally well defined
in the laboratory frame or in a fixed molecular frame. 

The most important consequence of this exchange of energy between electron
and core is that if the total energy $E$ is just slightly above the core energy
for the lowest core angular momenta allowed, we will find
points on this sphere, for which the electron energy $E_e$ is positive
and others for which it is negative. 
The positive electron energy region is thus given by $E>E_N=B N^2$,
which using ${\vec J}^2= \left({\vec L} + {\vec N}\right)^2$ gives
\begin{equation}
BN^2=B\left(-L\cos\alpha+\sqrt{J^2-L^2\sin^2\alpha}\right)^2<E
\end{equation}
This condition depends only on the angle $\alpha$ between ${\vec L}$
and the $\vec N$ axis, so that
the positive electron energy region is a cap about the positive $\vec N$-
axis.

Therefore we can describe a scattering
process by initiating the map with a point in the positive
electron energy region and
terminating it whenever we hit this region again. 
Indeed the restriction of the MQDT map to positive electronic energy
has been identified \cite{parabolic} with the Jung scattering map 
\cite{JungMap1,JungMap2}.
Ionization processes would correspond
to starting at some internal point and stopping when the
positive energy region is hit for the first time.

\section{Numerical Results}
\label{sec:NumericalResults}

\begin{figure*}
\TwoSides{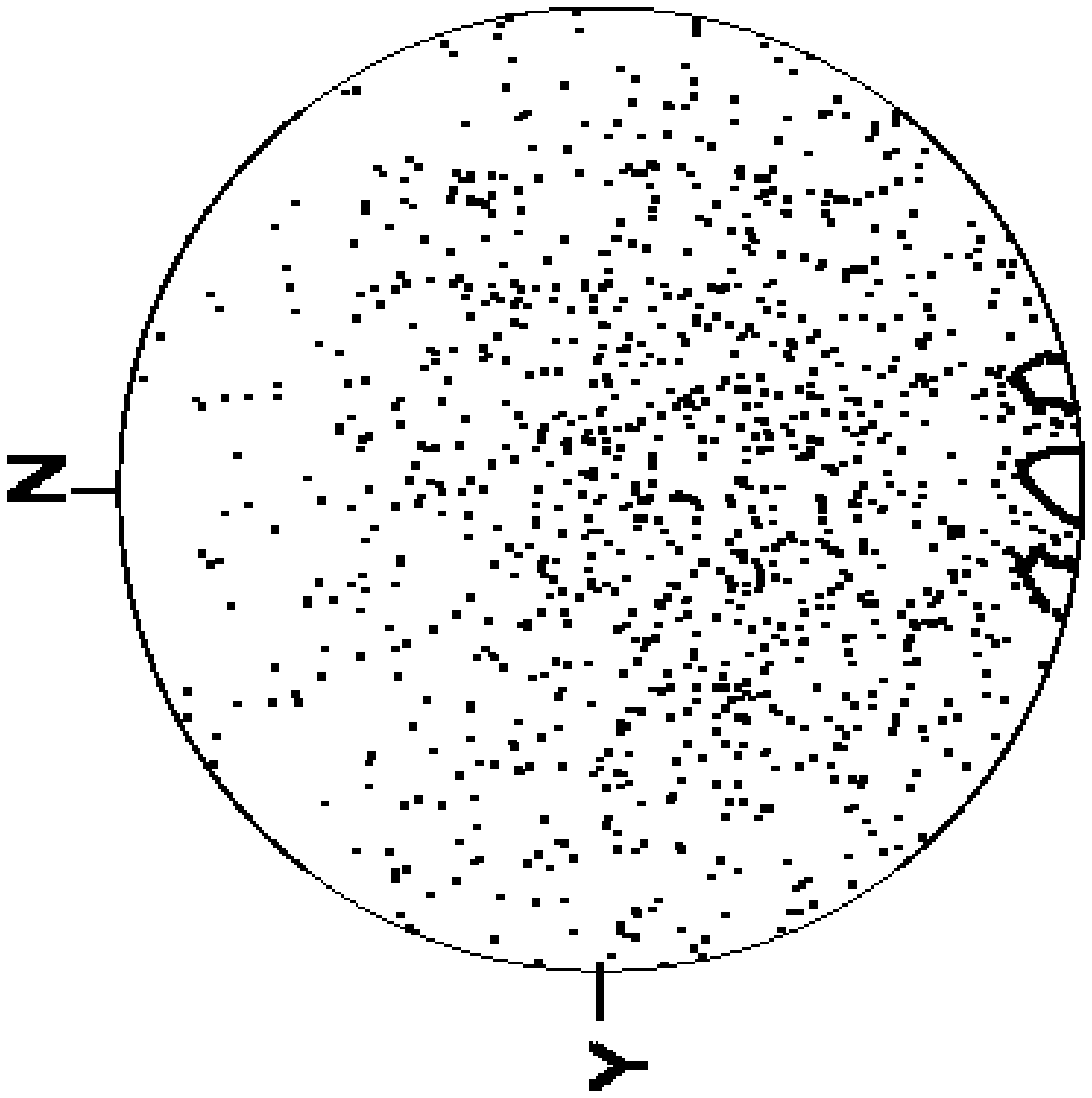}{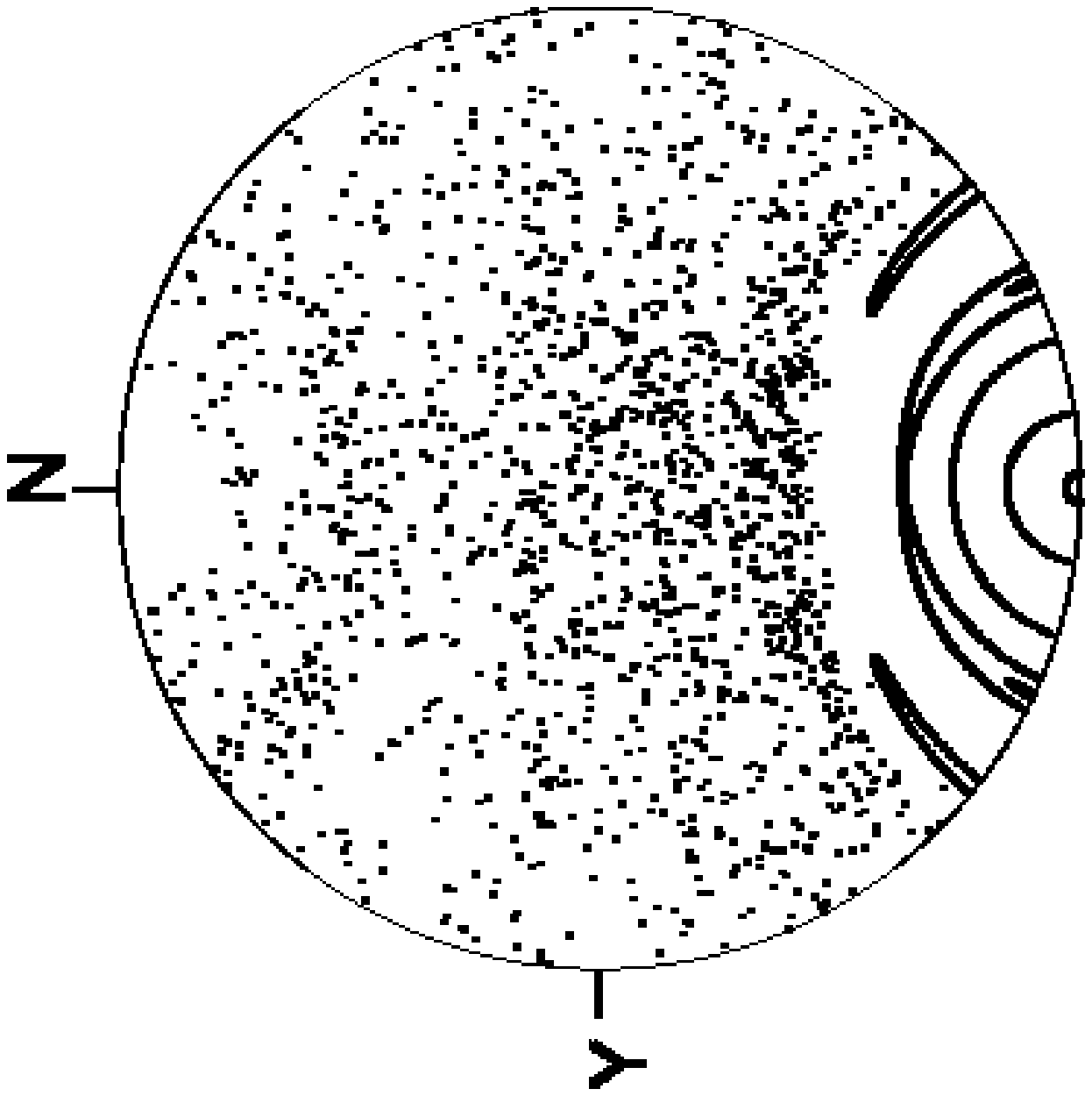}
\caption{Poincar\'e surfaces of section.
Stereographic projection from pole ($\vec P$).
The positive electron energy region is a small (white) cap
around positive $\vec N$, with angular radius $\alpha_0=0.075\pi$.
Left: strong coupling ($K=1$).
Right: small coupling ($K=0.1$).}
\label{fig:Poincare}
\end{figure*}

For the part of the sphere corresponding to negative electron energy,
one observes the usual KAM scenario as the coupling constant $K$ increases.
The size of the ordered regions decreases as $K$ grows.
To get an idea of the general KAM structure in Fig.~\ref{fig:Poincare}
we show views of the map at total energies $E$
just above the ionization energy for two different coupling constants.
Note that, for strong coupling, elliptic regions are (nearly)
invisible. For the weaker constant
these regions are large at the back of the sphere  and display the
typical features of a twist map. These elliptic
regions
correspond to positive (total) energy bound states,which are
stable in the classical regime.
In quantum mechanics we will expect them to decay slowly,
and it will be interesting, in some future work, to study their decay law,
which will not obviously be exponential, due to the fractal structure of
islands surrounding the main elliptic islands.

We next proceed to analyze the time delays in scattering both for
strong and weak coupling.
Here we have to address the question whether we wish to look at the
real time or at the number of iterations of the map. The real time
is certainly more attractive at first sight, but
the excursions of the electron on a Rydberg orbit take times that
approach infinity as the
electron energy approaches zero from below. Not only do these
long times distort the entire picture but they are actually
also unrealistic because any electron will feel some perturbation on a
sufficiently long excursion. We shall therefore first consider
the delay in terms of iterations
of the map in detail and at the very end show the true time evolution.

\begin{figure}
\centering
\includegraphics*[width=\figwidth,angle=270]{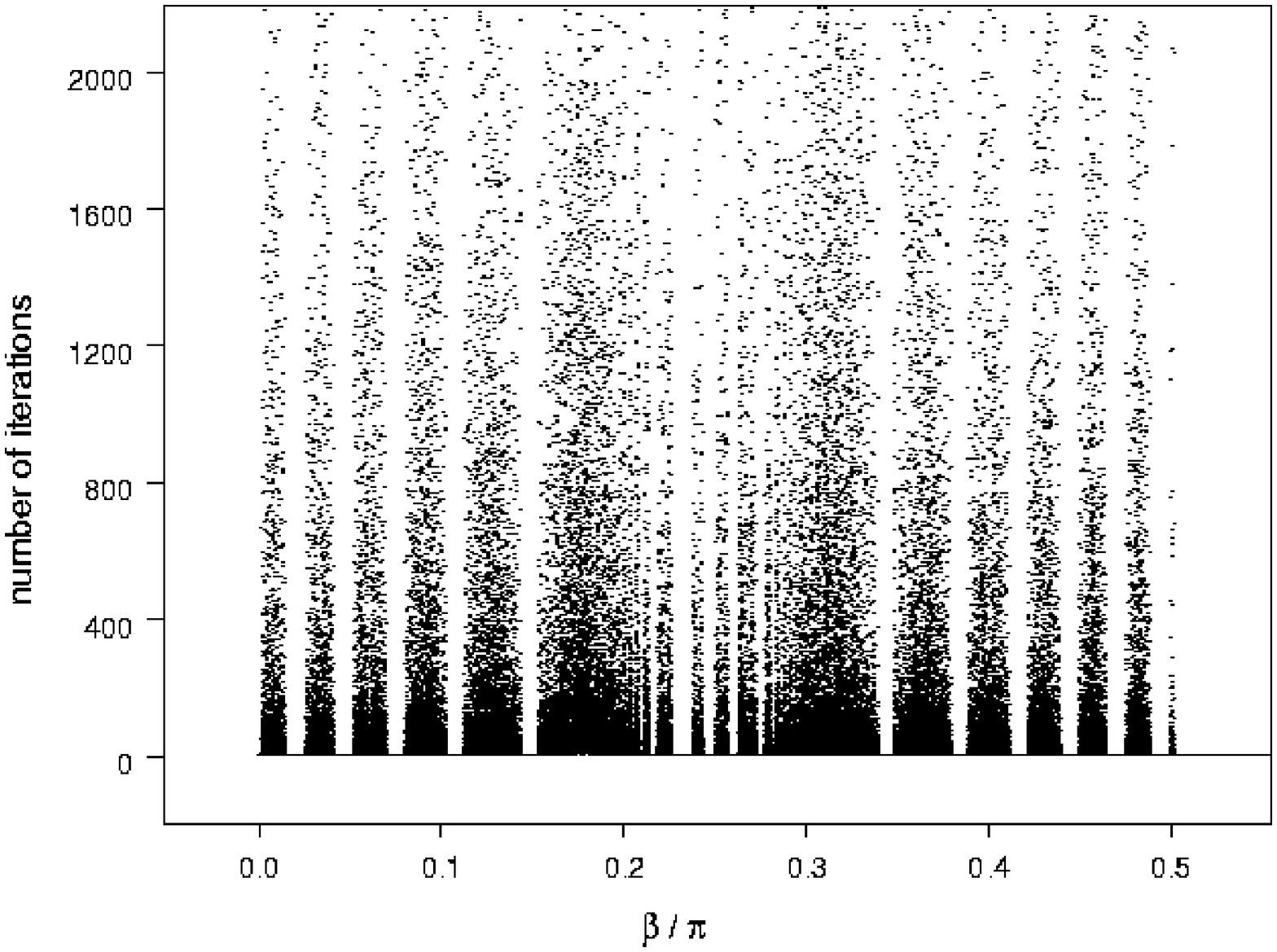}
\includegraphics*[width=\figwidth,angle=270]{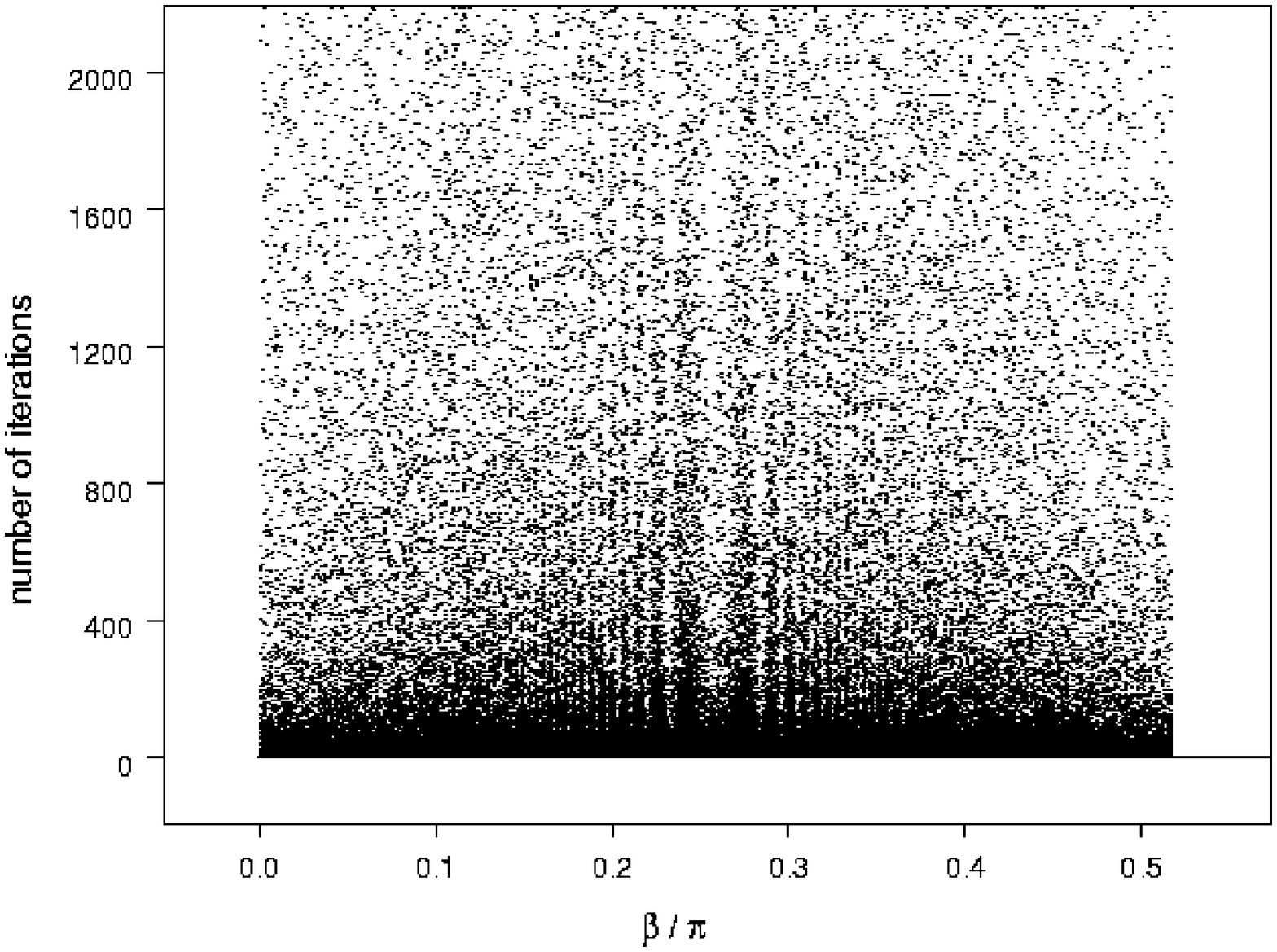}
\caption{Fractal structure of delays corresponding to artificial
ionizing thresholds. Top: artificial threshold $\alpha_0=0.5\pi$.
Bottom $\alpha_0=0.25\pi$. For actual zero energy threshold
the structure becomes too narrow to be seen.}
\label{fig:ScatteringTangle}
\end{figure}

The most common idea nowadays to study these scattering delays would be
to launch a set of trajectories on the circle corresponding to the
scattering threshold, and to display the fractal structure of these
delays as a function of the angle along this circle.
As said in the introduction we saw no structure this way.
To understand what is going on we use the following trick. Suppose
that for some reason (e.g. external electric field) the electron
is ionized for some negative electron energy $E_e$: this would
correspond to suppress far flung trajectories. We can then launch
a set of trajectories on this new threshold circle and study the new
delays. We have done this for the same parameters as for all other
cases in this paper (see the caption of table~\ref{tab:fit}), 
which correspond to a very narrow cap of angle $\alpha_0=0.075\pi$ 
of positive electron energies, but selecting artificial ionizing thresholds
electron energies which correspond to half sphere ($\alpha_0=0.5\pi$)
and quarter sphere ($\alpha_0=0.25\pi$).
The corresponding structure (delay versus angle $\beta$ on the ring
(see fig.~\ref{fig:schema})) is displayed on fig.~\ref{fig:ScatteringTangle}.
The half sphere case displays clearly the expected fractal delay structure
(we have checked that it is indeed fractal by expanding some parts
of this structure). For the narrower quarter sphere case the fractal
structure becomes very narrow, and for the true zero energy ionization
threshold it is no more visible.

We will thus now study the diffusion process, for which it will
turn out as we will see that these far flung trajectories instead of being
a drawback are an advantage because they are responsible
of the interesting structure we will see.

\begin{figure*}
\TwoSides{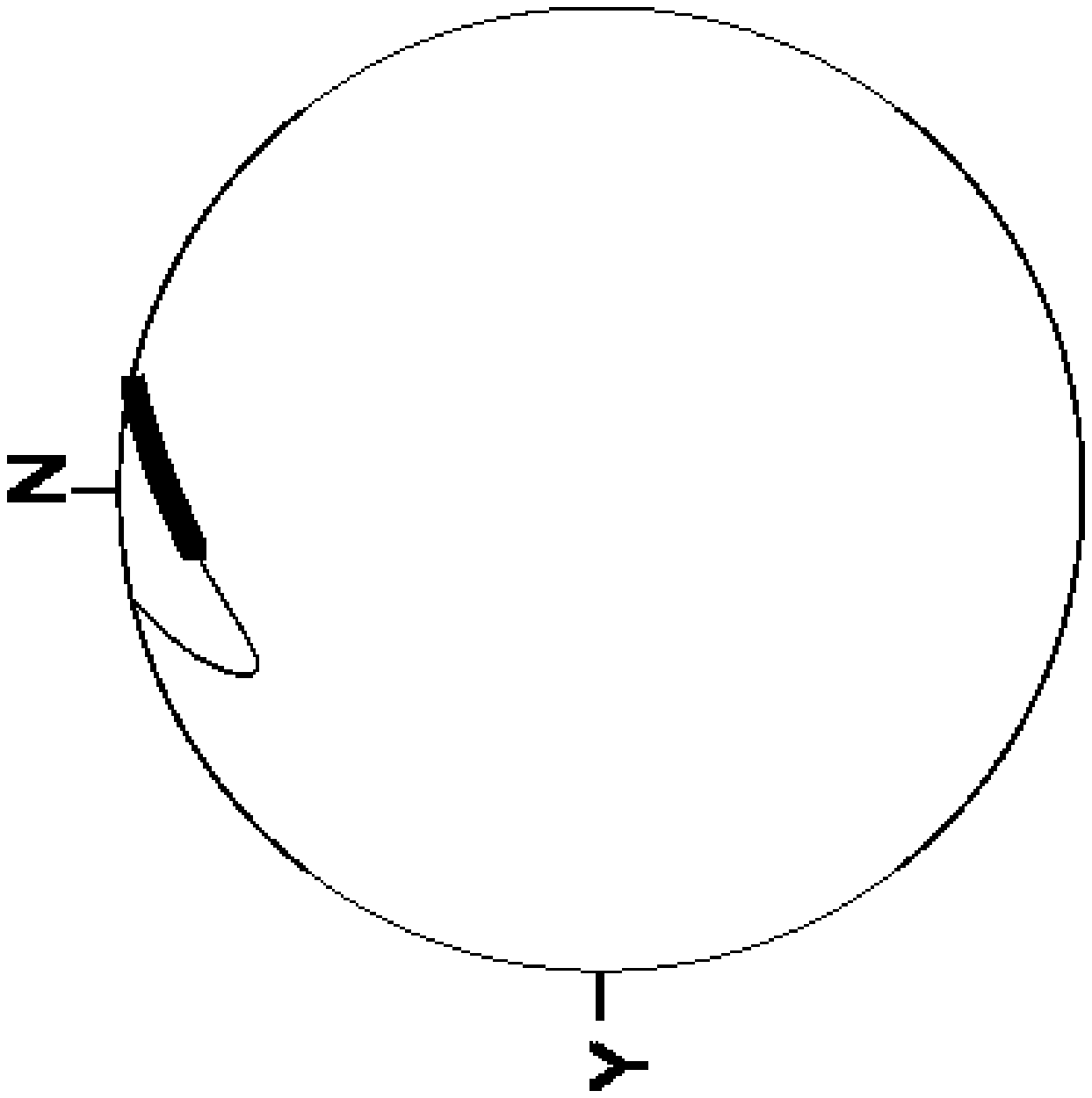}{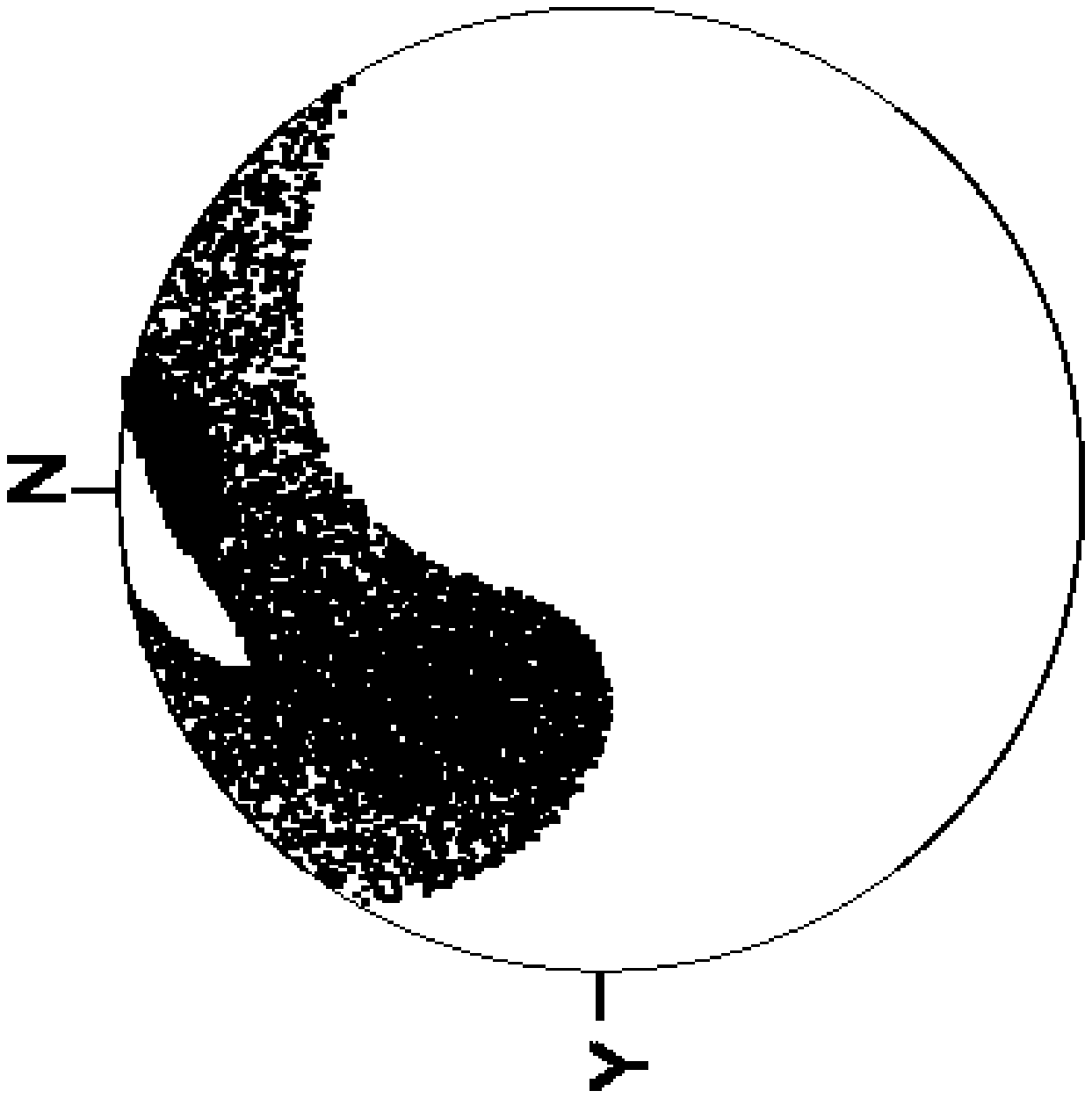}
\TwoSides{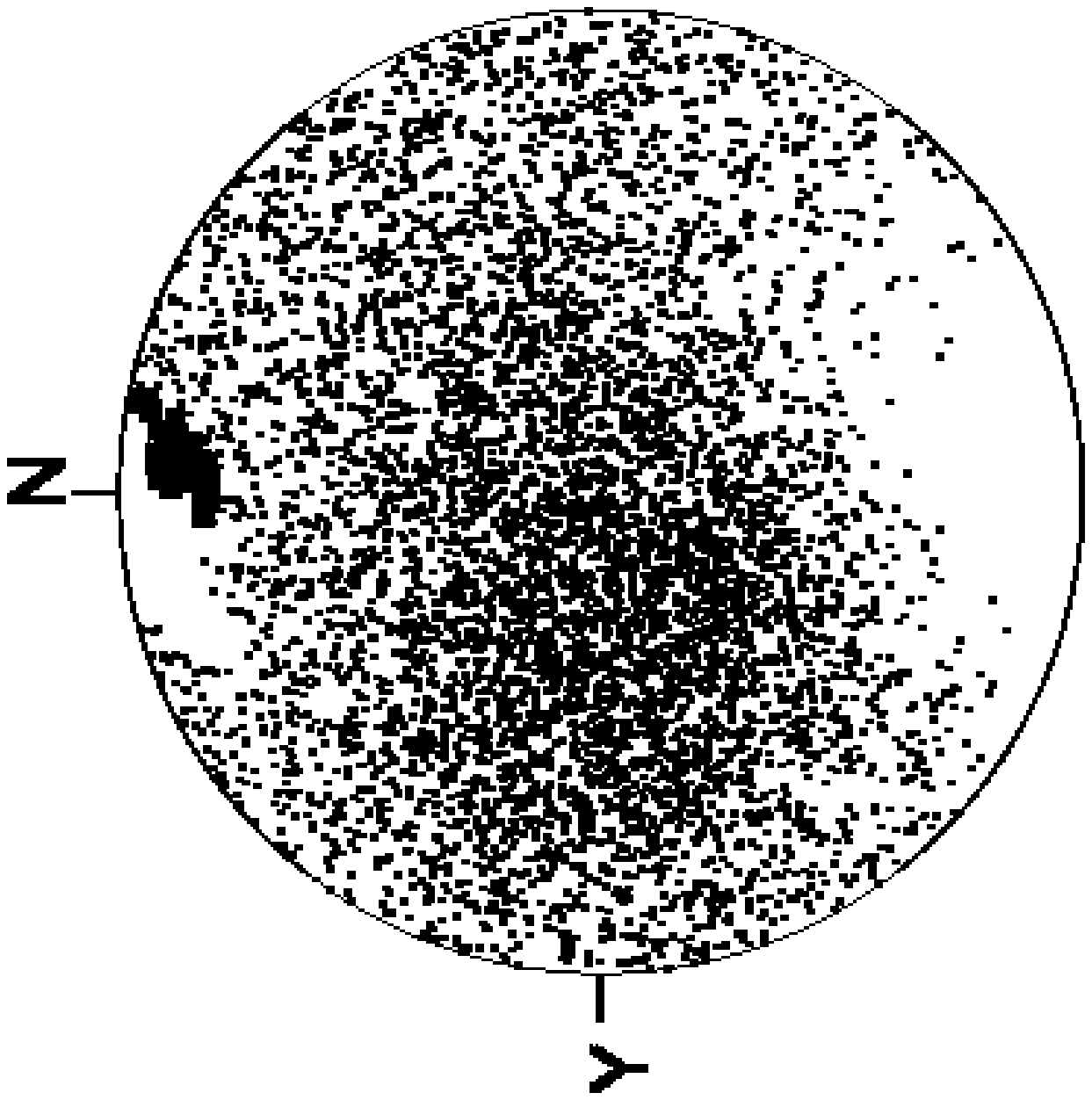}{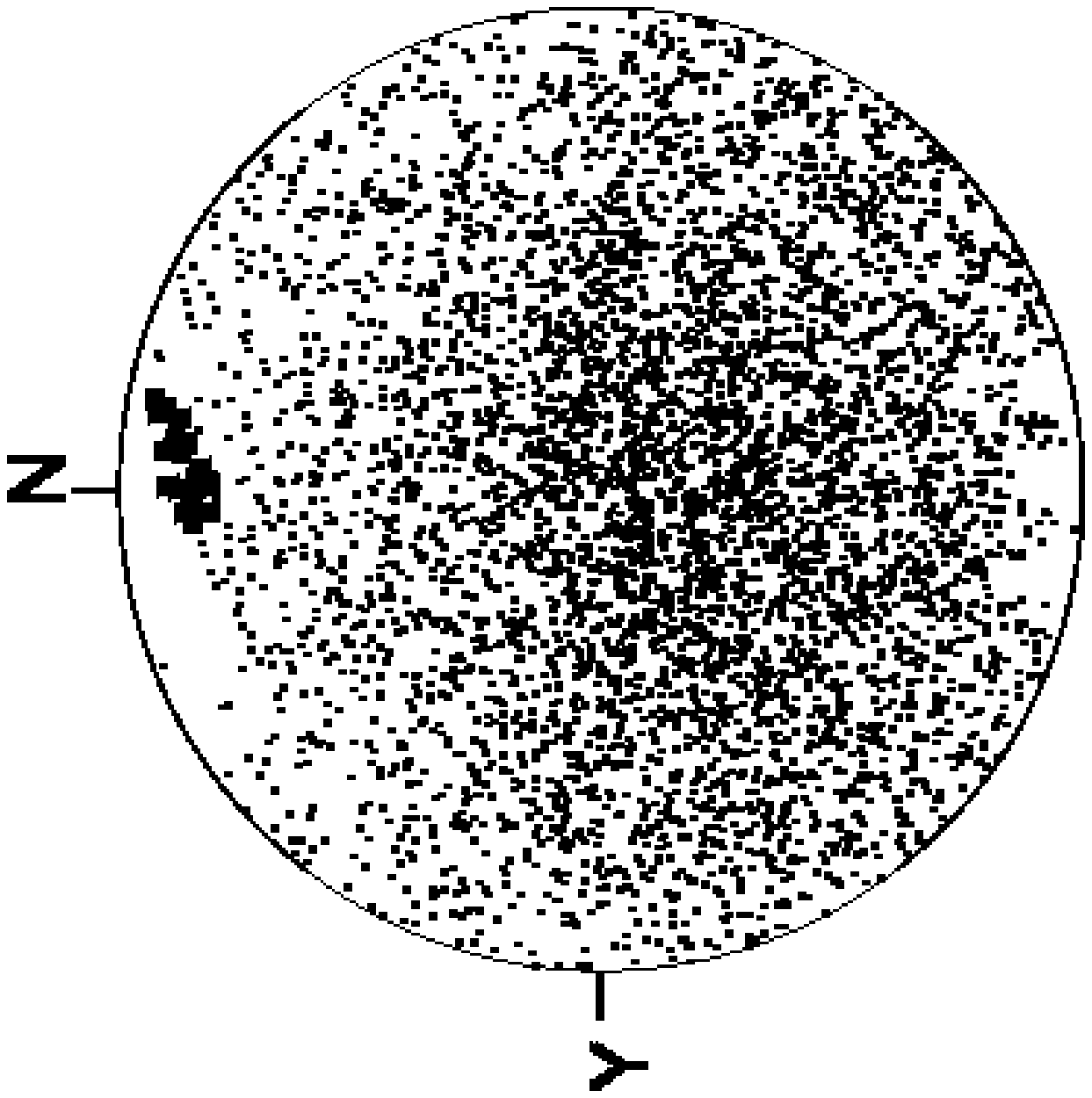}
\caption{Time evolution of a set of 20000 incoming particles
arranged on the circle that defines the scattering threshold:
large squares correspond to escaping particles.
Strong coupling: $K=-1$.
Top left: after 1 iteration (it): half the incoming particles 
are slowed down by the collision with the core and enter the sphere,
half are accelerated and escape. Top right: idem after 3 it.
Bottom left : after 10 it ($\sim t_c$ defined by Eq.~(\protect\ref{eq:tc})).
Bottom right: idem after 100 it.}
\label{fig:PoincareCircleStrong}
\end{figure*}

Let us first consider the strong coupling case $K=-1$:
Fig.~\ref{fig:PoincareCircleStrong} shows the time evolution of
a set of 20000 incoming particles
arranged on a ring concentric to the circle that defines the
scattering threshold. The sequence shows the original distribution
and its evolution after 1, 3, 10 and 100 iterations. Note that the
total number of particles diminishes due to scattered particles.
Thus the density diminishes but after a small number of
iterations this density tends to be very uniform, i.e. the system
has equilibrated. Note that the escaping particles are drawn as empty
squares that appear on the cap of the sphere that corresponds to
positive electron energies.

\begin{figure}
\centering
\includegraphics*[width=\figwidth,angle=270]{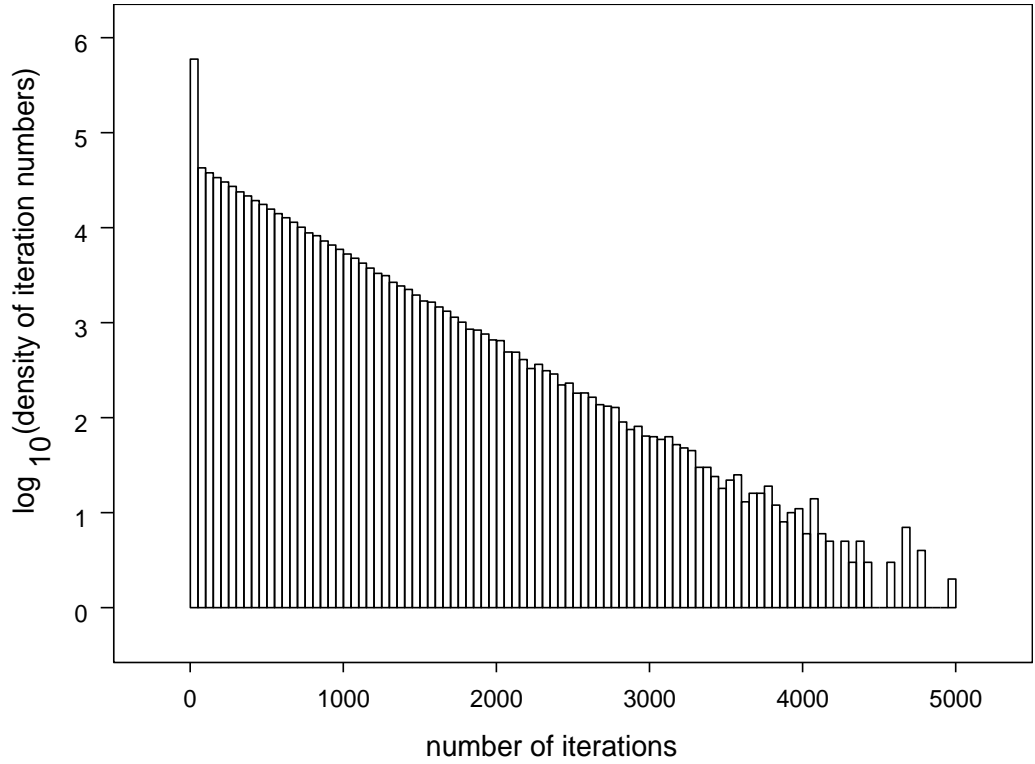}
\includegraphics*[width=\figwidth,angle=270]{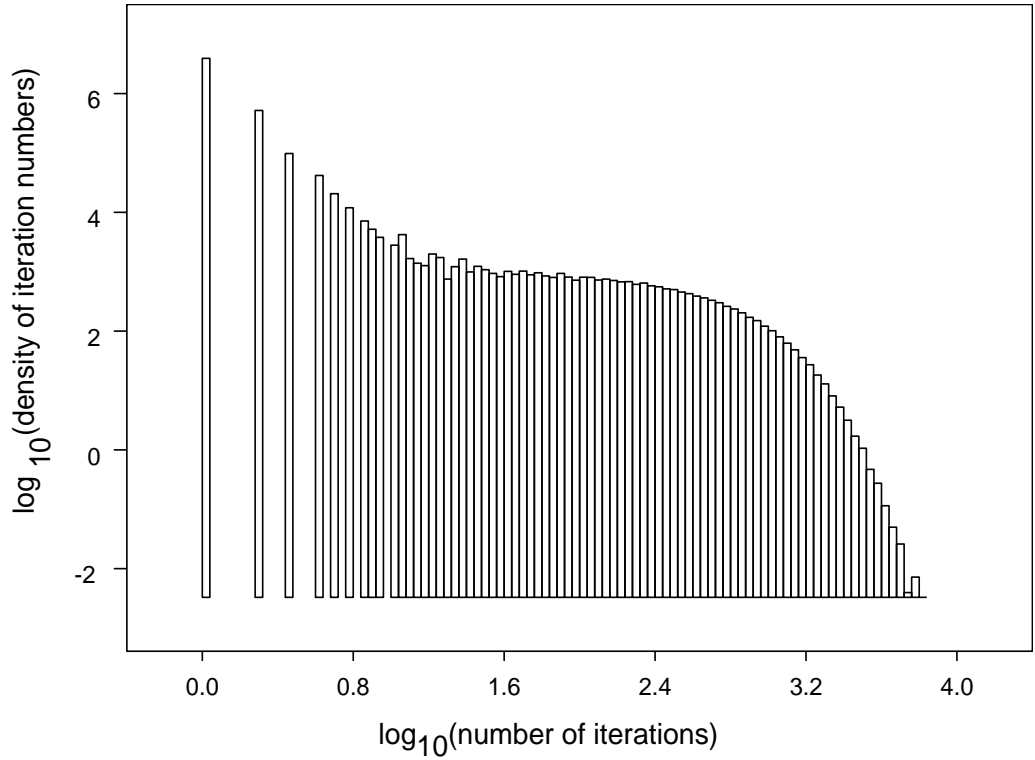}
\caption{Distribution of the number of iterations before escape.
Strong coupling $K=-1$.
Top: semi logarithmic plot, bottom: log log plot.}
\label{fig:HistoStrong}
\end{figure}

In Fig.~\ref{fig:HistoStrong} top, we show the number of particles
escaping
at a given iteration in a semi-logarithmic plot.
We see the straight line characteristic
of exponential decay dominating the figure, and an anomaly in the first
channel.
To get a better feeling for
this short time behavior, we next look at a log-log plot of the
same data in Fig.~\ref{fig:HistoStrong} bottom.
Here we see that the short time ($t\lesssim 10$)
behavior is a straight line {\it i.e.} a power-law.
Whereas the exponential behavior is standard for chaotic scattering,
the
power law for the transient is puzzling at first sight.

\begin{figure*}
\TwoSides{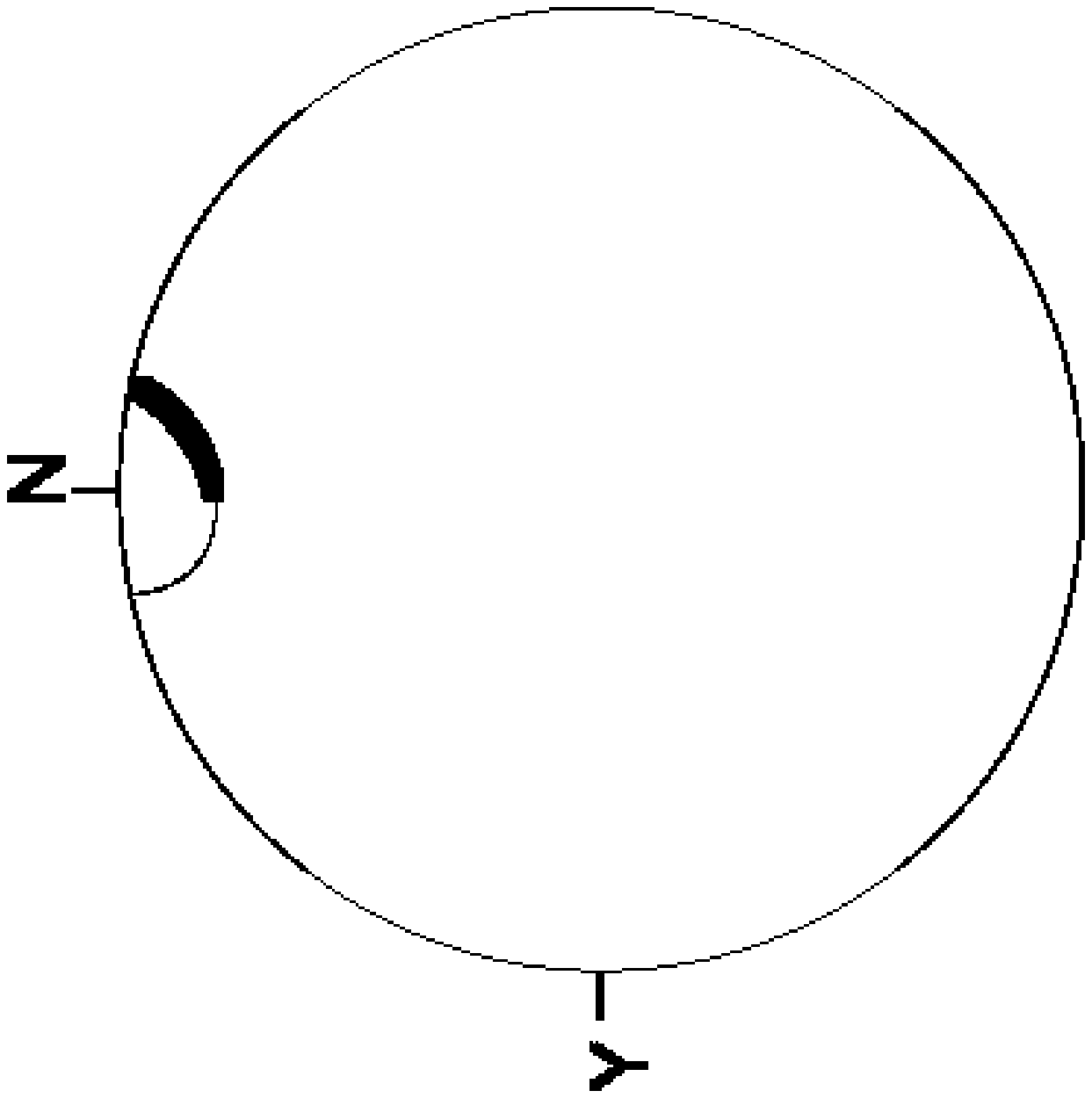}{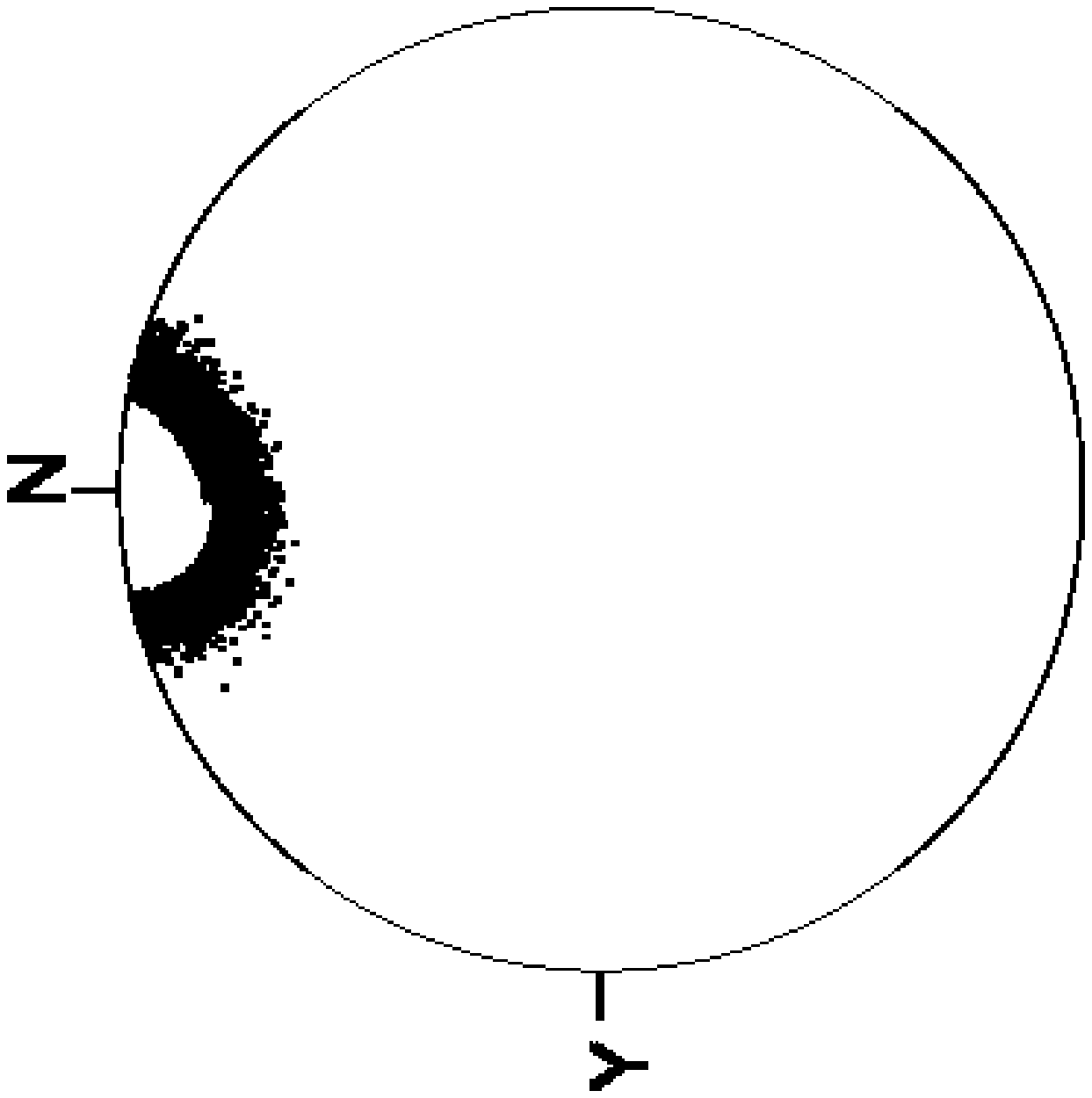}
\TwoSides{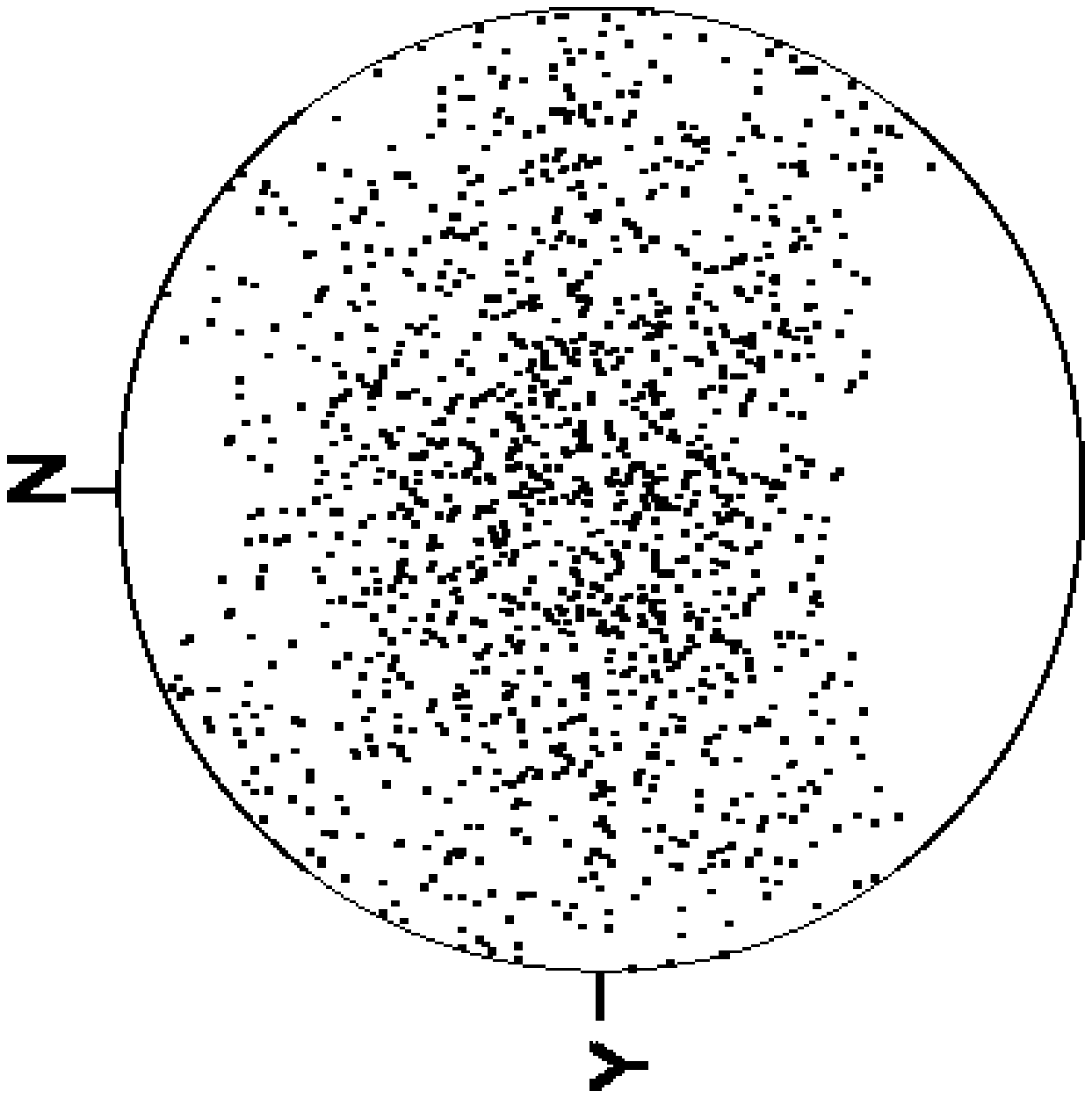}{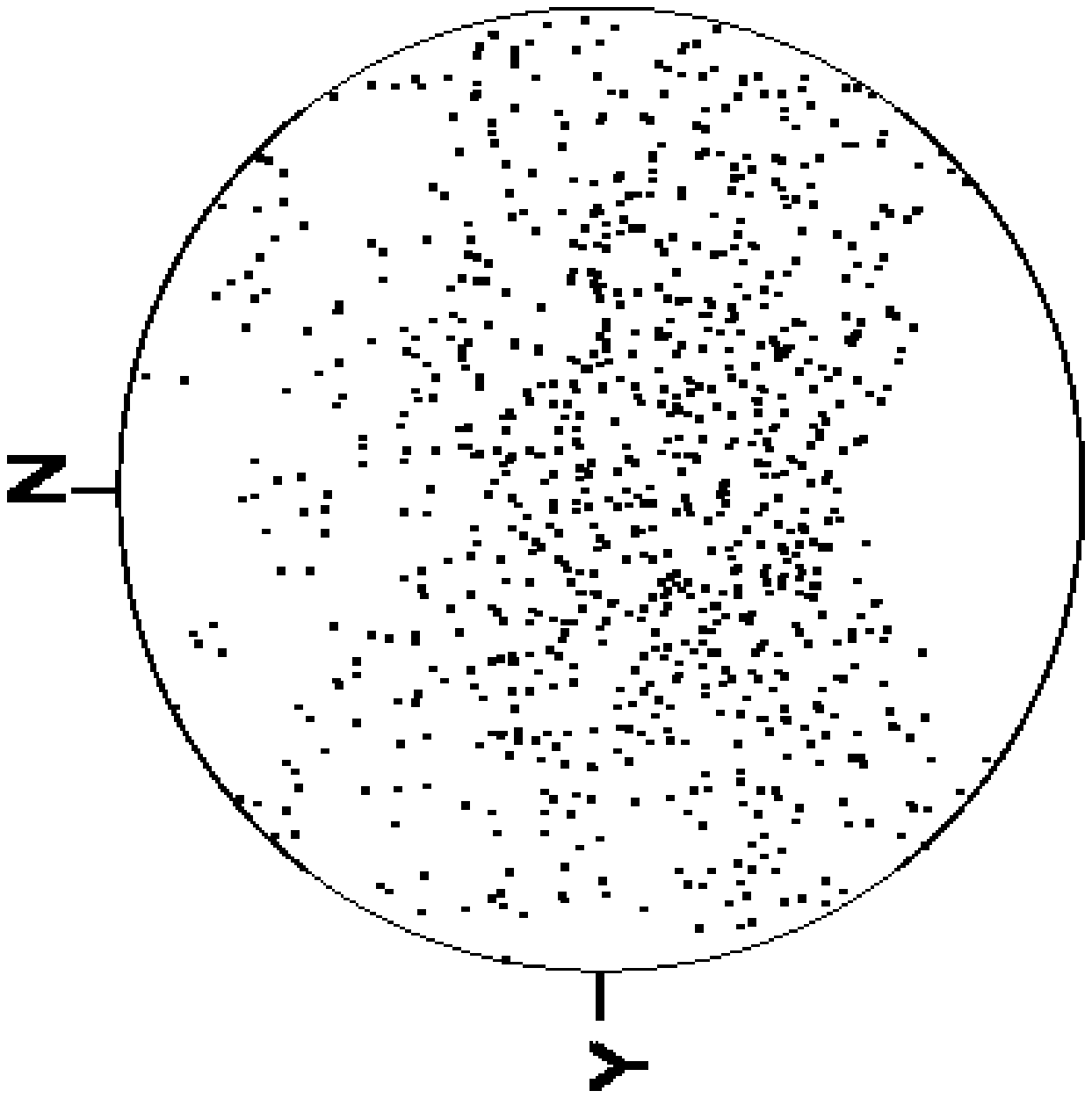}
\caption{Time evolution of a set of 20000 incoming particles
arranged on a ring concentric to the circle that defines the
scattering threshold: large squares correspond to escaping particles.
Weak coupling: $K=-0.1$.
Top left: after 1 iteration (it). Top right: idem after 10 it.
Bottom left: after 1000 it ($\sim t_c$ defined by 
Eq.~(\protect\ref{eq:tc}), see table~\protect\ref{tab:fit}).
Bottom right: idem after 10\,000 it.}
\label{fig:PoincareCircleWeak}
\end{figure*}

To gain better understanding we pass to the weak coupling case
given by $K=-0.1$.
Again we first look in Fig.~\ref{fig:PoincareCircleWeak}
at a sequence of pictures after
1, 5, 1000, 10\,000, iterations. Here we see that
the particles diffuse very slowly towards the back of the sphere and
they never penetrate the elliptic islands.
At very large numbers of iterations some get quite close, and thus
may penetrate the fractal tangle of small islands.

\begin{figure}
\centering
\includegraphics*[width=\figwidth,angle=270]{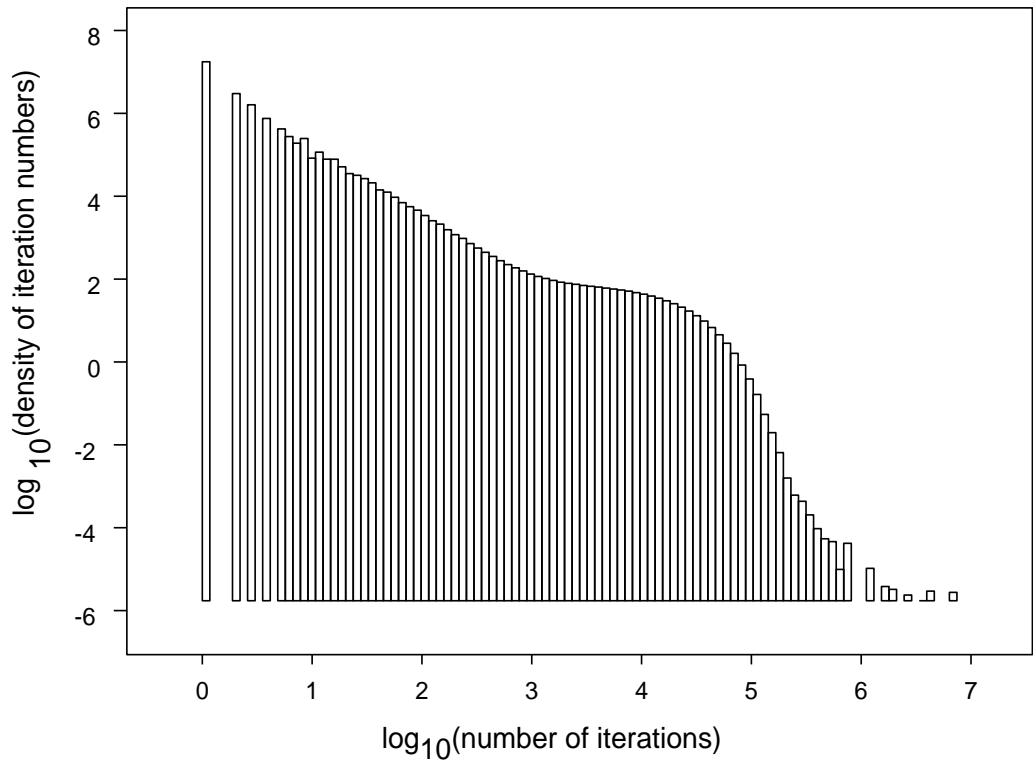}
\caption{Distribution of the number of iterations before escape.
Log log plot, weak coupling $K=-0.1$.}
\label{fig:LogLogWeak}
\end{figure}

To get a quantitative understanding of the observed behavior, we again
plot
the number of escaping particles versus iteration in a log-log plot
(fig.~\ref{fig:LogLogWeak}).
We see an initial power law as in strong coupling for now roughly 1000
channels. This power law can now be accurately fitted to determine
its power $-3/2$ (cf table~\ref{tab:fit}).
This result is compatible with the findings of fig.~\ref{fig:HistoStrong},
-bottom

\begin{figure}
\centering
\includegraphics*[width=\figwidth,angle=270]{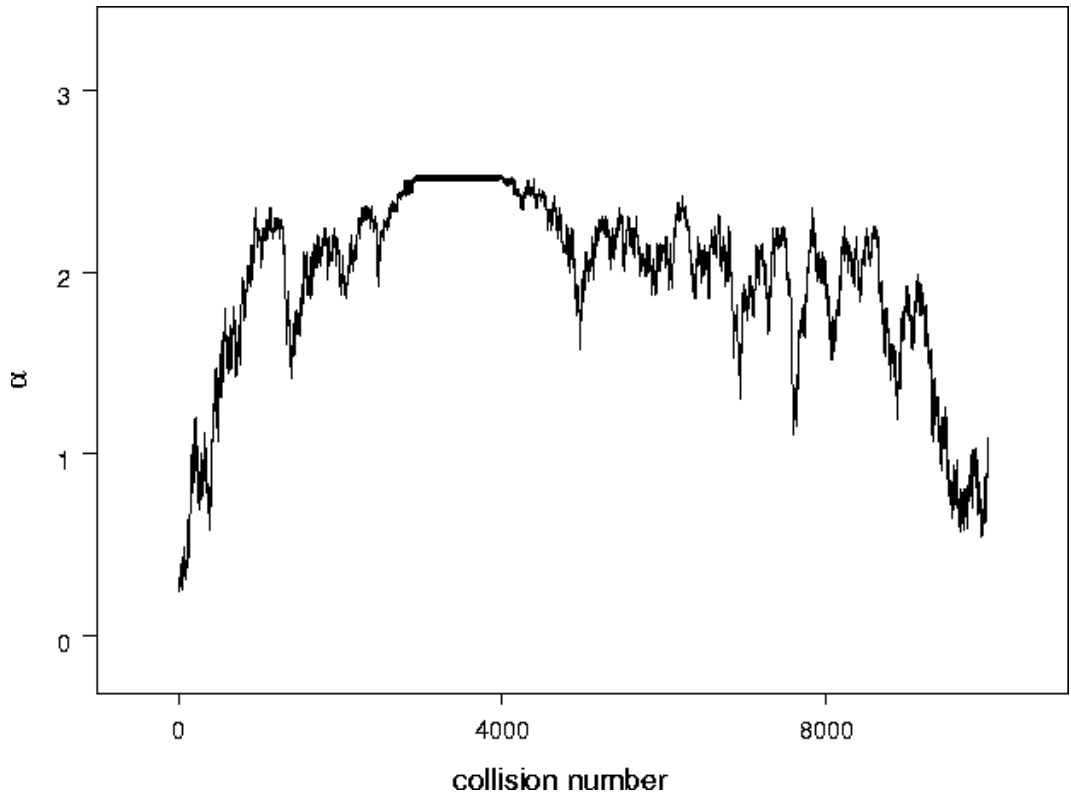}
\caption{Some particles are caught in the chaotic tangle:
for a number of iterations $\alpha$ remains nearly constant
at a value corresponding
to the chaotic tangle (cf Fig.~\protect\ref{fig:Poincare}).
Weak coupling $K=-0.1$.}
\label{fig:Trappee}
\end{figure}

Furthermore in this weak coupling case
an other distinct deviation from exponential behavior appears at very
long times ($t\gtrsim 2\,10^5$).
This last region is compatible with a power law between -1 and -3.
Let us treat first this second power law. This is what we expect
for particles caught in the chaotic tangle
\cite{PowerLawTangle1}\cite{PowerLawTangle2}. That this is indeed the
effect we are seeing, can easily be shown as follows.
In fig.~\ref{fig:Trappee} we see that some particles show
regions in their time-evolution plot where the angle $\alpha$
between ${\vec L}$ and $\vec N$ does not change over many
iterations indicating trapping in the fractal tangle around an
island which has this behavior according to fig.~\ref{fig:Poincare}.
We may actually subtract these iterations and make a new plot of delays
where these are absent as shown in Fig.~\ref{fig:ElimineTrappees}.
The final power law tail
disappears, but the lower times region is not affected.

\section{Diffusion Model}

\begin{figure}
\centering
\includegraphics*[width=\figwidth,angle=270]{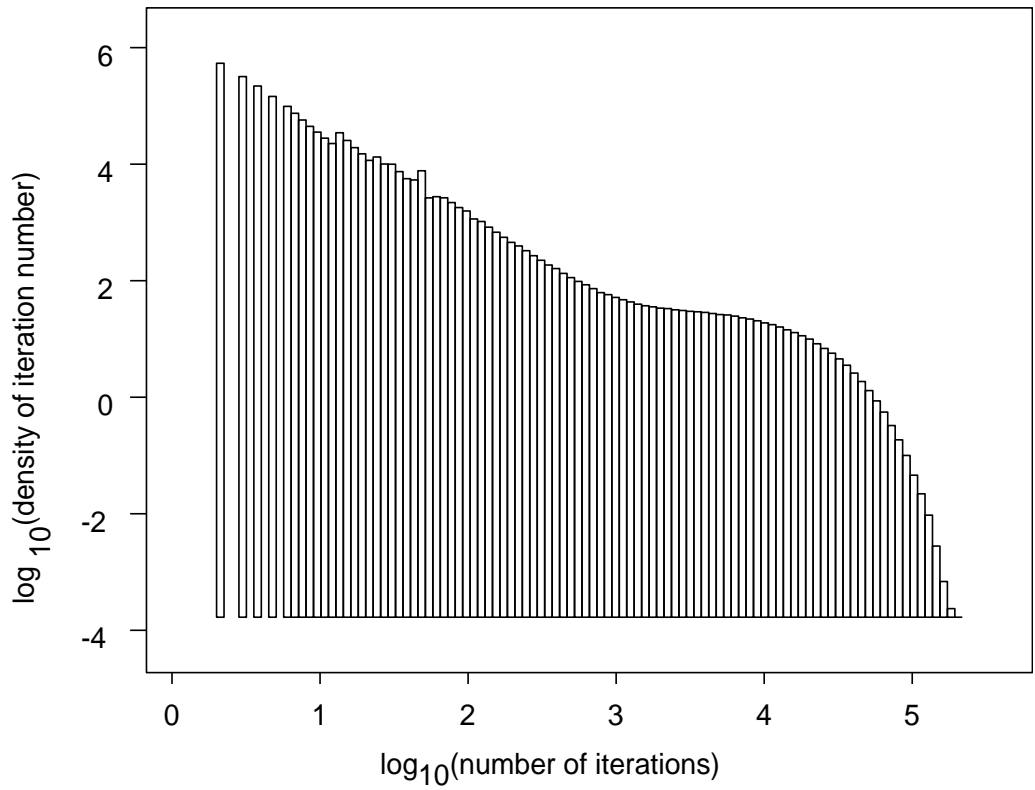}
\caption{Distribution of the number of iterations before escape,
eliminating the iterations caught in the chaotic tangle.
Weak coupling $K=-0.1$. Log log plot.}
\label{fig:ElimineTrappees}
\end{figure}

\begin{figure}
\centering
\includegraphics*[width=\figwidth,angle=270]{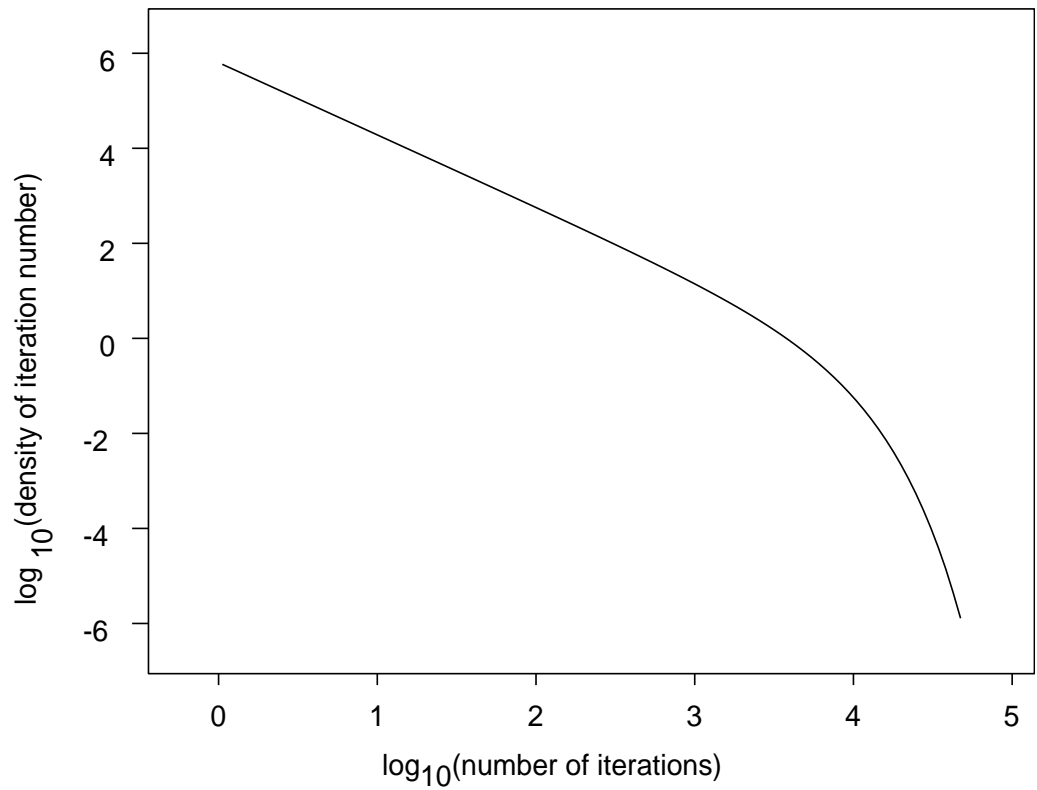}
\caption{Theoretical power law for a random walk in 1 dimension
Eq.~(\protect\ref{eq:random1D}). Log log plot.}
\label{fig:TheorieRandom1d}
\end{figure}

Let us focus now on the interpretation of this first power law,
which is the most interesting part of our results. The law
\begin{equation}
A t^{-3/2} \exp(-t/T)\label{eq:random1D}
\end{equation}
corresponds, as is well-known, to the probability that a 1D
random walk starting from the origin in an interval
of length $L\sim\sqrt T$ returns to the origin
for the first time at time $t$\cite{RandomWalk}.
We could readily explain the connection to the Rydberg
model by the following argument:
Our map consists of two parts: during the long Rydberg excursion
of the electron the core rotates around the $\vec N$ axis very often
and we have to take the total
rotation modulo $2\pi$ which makes for a reasonable randomizer of the
rotation angle $\beta$. Yet the kick $\delta\phi= K\cos\theta$
produced in the orthogonal direction (around $\vec P$) depends on the final
value of this angle $\beta$. In particular it is towards the front
or back of the sphere depending whether this $\vec N$ rotation
left it on the upper $z>0$ or lower $z<0$ part of the sphere.
It thus becomes a random walk towards the front or back of the sphere.
This could explain our findings.
However there is a distinct qualitative difference between this
1D random walk law and our Rydberg results. It is displayed
by comparing the previous fig.~\ref{fig:ElimineTrappees} with the
theoretical fig.~\ref{fig:TheorieRandom1d}.
Indeed, the law (\ref{eq:random1D})
always leads to a value that is lower than
the corresponding power law $At^{-3/2}$, whereas
in our Rydberg case there is a long, nearly horizontal, shoulder
between the power law and the exponential decay.
Mathematically the two histograms,
(\ref{fig:LogLogWeak}) and (\ref{fig:ElimineTrappees})
for the Rydberg case can be fitted by a formula
\begin{equation}
\left(A_1 t^{-\alpha_1} + A_2\right) \exp(-t/T)+A_3 t^{-\alpha_3}
\label{eq:loiRydberg}
\end{equation}
with $A_3=0$ for the latter case.
Parameters are given in table~\ref{tab:fit}.
There are negligible residuals: the sluggish transition
between the two parts is an effect of
non additivity in a log-log plot, and there is no statistically
significant possibility to fit two independent time constants
for $A_1$ and $A_2$.
This introduces a second characteristic time,
\begin{equation}
t_c=(A_1/A_2)^{1/\alpha_1},
\label{eq:tc}
\end{equation}
which is the transition time between power
law and exponential regimes, besides the
time constant of the exponential decay $T$, which is much longer.
The time $t_c$ has no equivalent in the 1D random walk model.
In both strong and weak coupling cases $t_c$
corresponds roughly according to
figs.~\ref{fig:PoincareCircleStrong}
and \ref{fig:PoincareCircleWeak}
to the time needed for the particles to fill two thirds of the sphere.

\begin{figure}
\centering
\includegraphics*[width=\figwidth,angle=270]{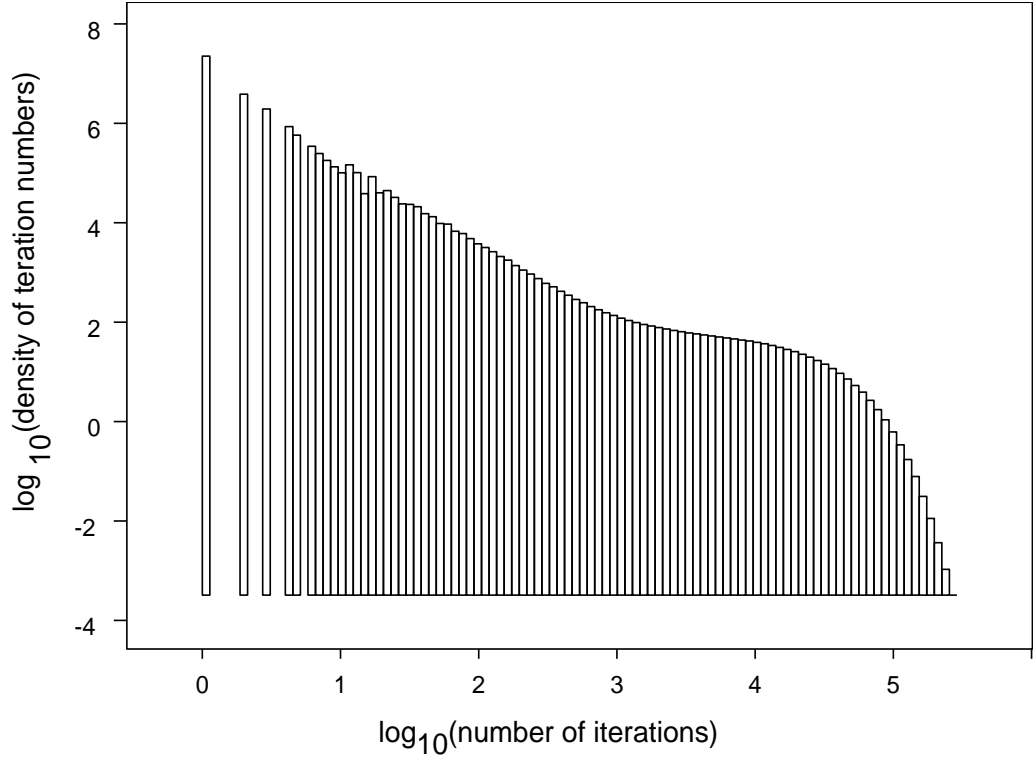}
\caption{Distribution of the number of iterations before escape,
replacing the true $\delta\beta$ by a random one,
while keeping the true Rydberg $\delta\phi+\delta\phi^\prime$.
The long time tail has disappeared, but the exponential
regime remains in addition to the power law.
Weak coupling $K=-0.1$.}
\label{fig:RandomBeta}
\end{figure}

\begin{figure}
\centering
\includegraphics*[width=\spherewidth,angle=270]{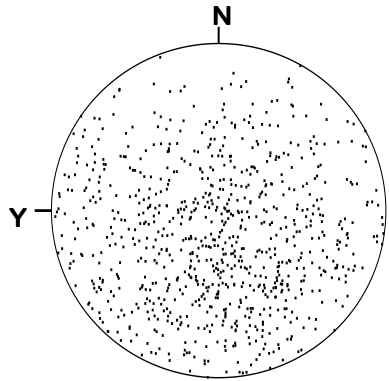}
\caption{Time evolution of a set of 20000 incoming particles
arranged on a ring concentric to the circle that defines the
scattering threshold, with random $\delta\beta$ and true
Rydberg $\delta\phi+\delta\phi^\prime$. Picture after 10\,000
iterations. With respect to the last of 
figs.~\protect\ref{fig:PoincareCircleWeak}
the blank part at the bottom of the sphere which corresponds to
the regular region has disappeared.
Weak coupling: $K=-0.1$.}
\label{fig:PoincareCircleRandomBeta}
\end{figure}

To check these ideas we replace the true rotation $\delta\beta$
by a random one, equidistributed on $[0,2\pi)$, while keeping the
exact Rydberg formula for the collisional kick
$\delta\phi+\delta\phi^\prime$
and resulting $\delta\alpha$, which is the significant parameter
to monitor distance to ionization.
The result (fig.~\ref{fig:RandomBeta} and table~\ref{tab:fit}) is
virtually indistinguishable from the case where trapped parts
of trajectories are eliminated (fig.~\ref{fig:ElimineTrappees}).
Since in this case the regular part at the back of the sphere
has disappeared (see fig.~\ref{fig:PoincareCircleRandomBeta}),
this result first corroborates our previous conclusion that the
long time tail is due to trajectories trapped in the chaotic tangle.
Second it shows that the model of random $\beta$ is correct
but that the right Rydberg value of the kick
$\delta\phi+\delta\phi^\prime$
introduces a qualitative difference with respect to the simple model
of a constant amplitude kick.
To gain further insight, since the formula for
$\delta\phi+\delta\phi^\prime$
is too complex to enable to visualize easily its effects, we
compute a first order development of $\delta\alpha$
as a function of the small parameter $K$:
\begin{equation}
\delta\alpha=
\left(1-\frac{L}{\sqrt{J^2-L^2\sin^2\alpha}}\cos\alpha\right)^{-1}
\half K \sin\alpha \sin2\beta
\end{equation}
Neglecting the recoil $\delta\phi^\prime$, which is valid if $L \ll J$,
it simplifies further as:
\begin{equation}
\delta\alpha=\half K \sin\alpha \sin2\beta
\label{eq:DeltaAlphaOrdre1}
\end{equation}

\begin{figure}
\centering
\includegraphics*[width=\figwidth,angle=270]{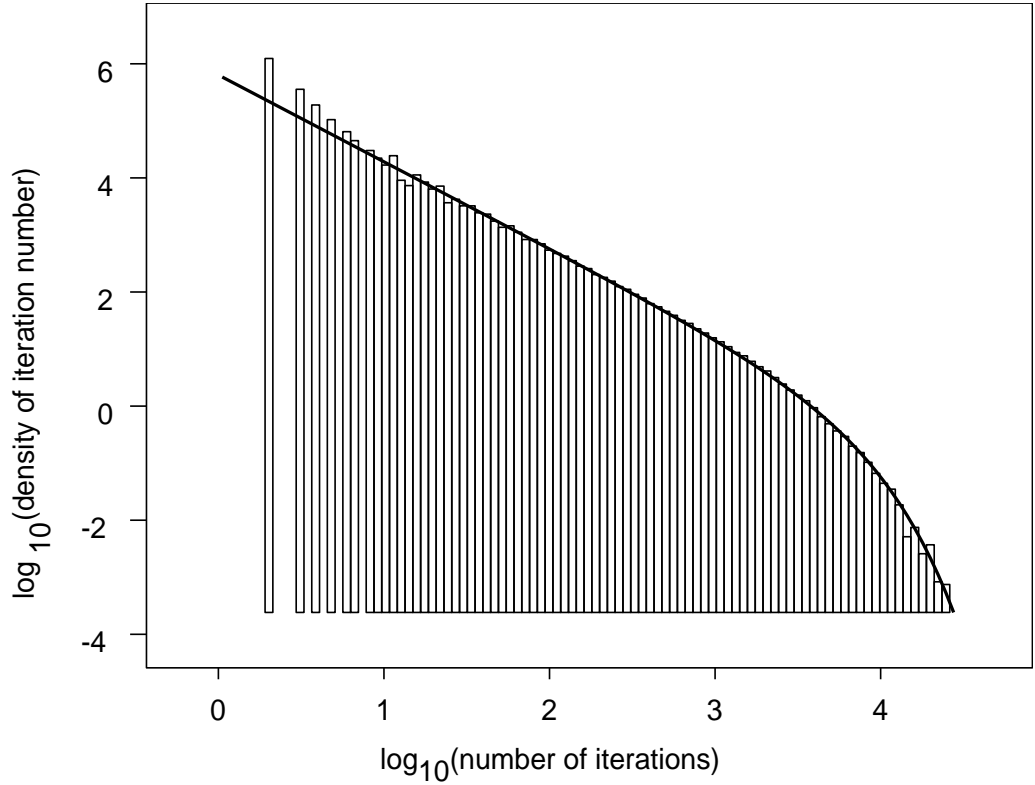}
\caption{Distribution of the number of iterations before escape,
replacing the true $\delta\beta$ by a random one, and the true
$\delta\alpha$ by the first order (in $K$) value
of Eq.~(\protect\ref{eq:DeltaAlphaOrdre1}).
Heavy line: fit by Eq.~(\protect\ref{eq:random1D}), with
parameters given in table~\protect\ref{tab:fit}.
Weak coupling $K=-0.1$.}
\label{fig:RandomOrdre1}
\end{figure}

\begin{figure}
\centering
\includegraphics*[width=\figwidth,angle=270]{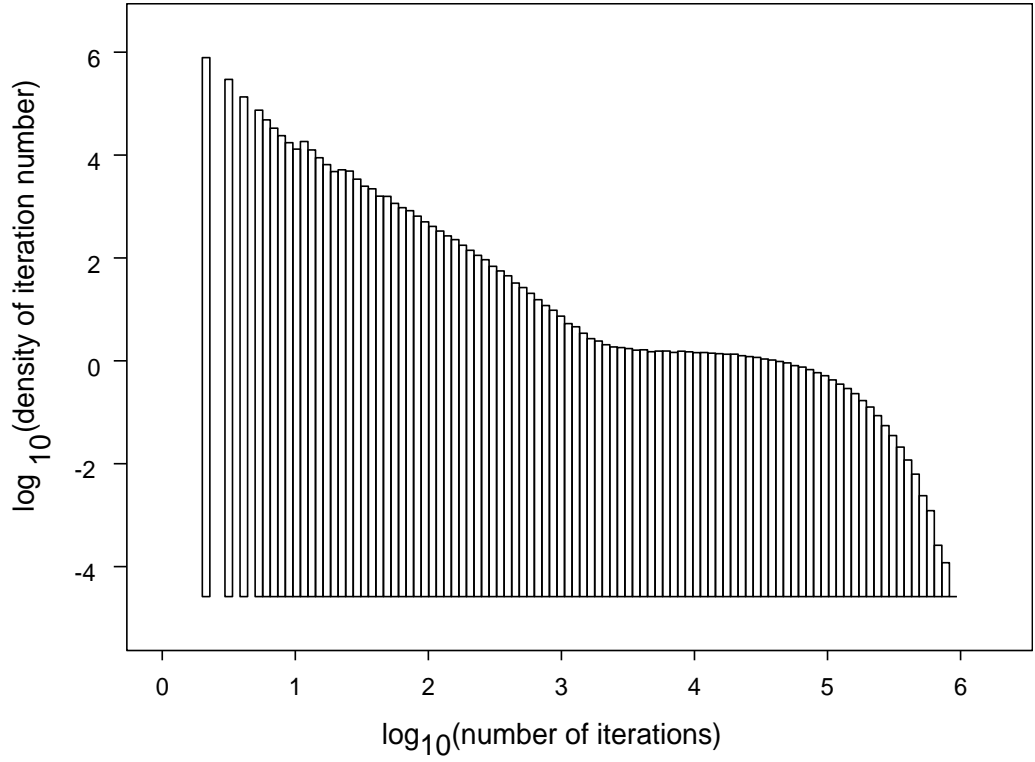}
\caption{Distribution of the number of iterations before escape,
replacing the true $\delta\beta$ by a random one, and the true
$\delta\alpha$ by the second order (in $K$) value
of Eq.~(\protect\ref{eq:DeltaAlphaOrdre2}).
Weak coupling $K=-0.1$.}
\label{fig:RandomOrdre2}
\end{figure}

With any of these first order formulas, the resulting histogram
(fig.~\ref{fig:RandomOrdre1} and table~\ref{tab:fit}) corresponds
to the 1D random walk prediction of Eq.~(\ref{eq:random1D})
and fig.~\ref{fig:TheorieRandom1d}, with no exponential shoulder.
What is characteristic in these equations is that for an equipartition
of $\beta$ on $[0,2\pi)$ the average value of $\delta\alpha$ is zero.
The exact value of the law, especially the way it varies with $\alpha$
is of little importance. This is expected since a variation of
$|\delta\alpha|$ with $\alpha$ can be compensated for by a
renormalisation of $\alpha$ to keep its variation constant on average.
The existence of the exponential shoulder is thus related to the
fact that the true $\delta\alpha$ is not zero on average. This
is shown by writing $\delta\alpha$ up to second order in $K$
(neglecting $\delta\phi^\prime$ to simplify):
\begin{equation}
\delta\alpha=\half K \sin\alpha \sin2\beta
+\quart K^2 \sin2\alpha \sin^4\beta.
\label{eq:DeltaAlphaOrdre2}
\end{equation}
The resulting histogram (fig.~\ref{fig:RandomOrdre2}) displays indeed
the exponential shoulder.
The explanation is as follows.
The average value of $\delta\alpha$ is positive for the front part
of the sphere ($0<\alpha<\pi/2$), thus pushing it backwards towards
the middle (increasing $\alpha$), and conversely negative for the back
part of the sphere, pushing it also towards the middle.
This creates a kind of trap in the middle of the sphere,
and the exponential law results from the need to escape that trap.

\begin{figure}
\centering
\includegraphics*[width=\figwidth,angle=270]{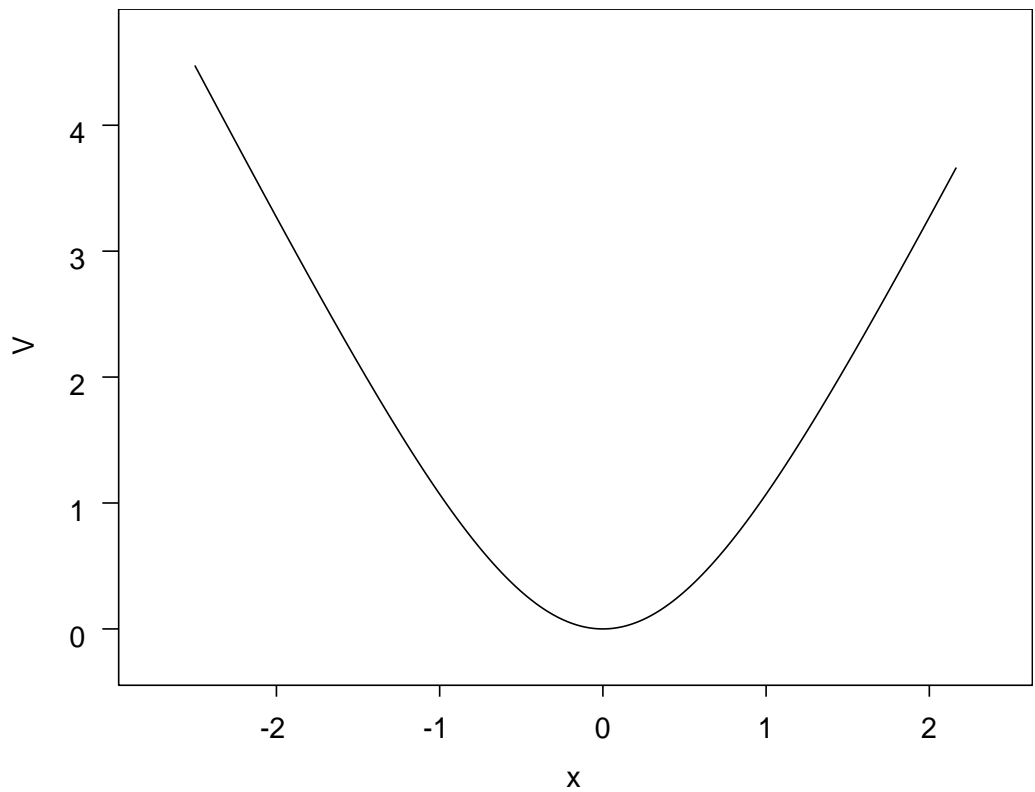}
\caption{Effective potential defined in
Eq.~(\protect\ref{eq:Potentiel}). Ordinate unit is $kT$.}
\label{fig:Potentiel}
\end{figure}

To justify this explanation let's put it quantitatively.
We first renormalize $\alpha$ to have
constant average $|\delta\alpha|$ by redefining a new coordinate
\begin{equation}
x=\half\ln\frac{1+\cos\alpha}{1-\cos\alpha}
\end{equation}
which varies between $-\infty$ and $+\infty$ when
$\alpha$ varies between $\pi$ (back) and $0$ (front of the sphere).
$x$ has been computed so that to first order $\delta x$ is
independent of $x$, and then to second order in $K$
\begin{eqnarray}
\delta x&=&-\half K\sin2\beta-
\half K^2 \cos\alpha\left(\sin^4\beta-\quart\sin^2 2\beta\right)
\nonumber\\
&=&-\half K\sin2\beta-
\half K^2\tanh x\left(\sin^4\beta-\quart\sin^2 2\beta\right)
\end{eqnarray}
We obtain thus a random motion with differing average amplitude
in the two directions. This is equivalent to a thermal motion
within a potential $V(x)$. The average step is $\pm K/2\pi$ and the
difference between the two directions comes from a Boltzmann factor:
\begin{equation}
\exp\left(-{V'(x)\times K/\pi}\right)
\simeq 1-{V'(x)\times K/\pi}
=\frac{\langle\delta x^+\rangle}{\langle\delta x^-\rangle}
=1-\frac{\pi}{4} K \tanh x,
\end{equation}
where we have normalized temperature and potential so that $kT=1$. Thus
\begin{equation}
{V(x)}=\frac{\pi^2}{4} \ln\cosh x
\label{eq:Potentiel}
\end{equation}
This potential is plotted in fig.~\ref{fig:Potentiel}.
The potential curve is limited on the right at the actual value
of $x$ corresponding to the
ring of ionization in all preceding figures.
The depth of the trap is $3.6$, justifying a long time
exponential to escape it.

In order to compare the results given here with
the times reported in table~\ref{tab:fit}, one should note
the following: these times are given
in number of iterations. However, at each iteration, $x$ changes
by an amount of order $K$, so that the
natural diffusive time scale, over which
$x$ varies by a quantity of order one,
is of order $K^{-2}$. We therefore expect
the times given in table~\ref{tab:fit} to contain an overall
factor of order $K^{-2}$ apart from the
first passage probability.
That this is indeed so, is verified
in the case $K=0.2$, for which
all times are approximately four times
shorter than for $K=0.1$.

The physical origin of this potential is obvious.
The electron can both gain or loose energy at
each collision with the core.
The most probable situation is equipartition of energy, i.e. the
middle of the sphere, and ionization
which needs nearly all energy in the electron is a rather improbable
situation. This potential takes thus into account the amount of
available phase space.

We now give a theory of this phenomenon.
We wish to solve the following problem:
if we introduce a particle at the point $a$
and this particle diffuses under the influence
of a binding potential $V(x)$, what is the
probability that it returns
to $a$ for the first time at time $t$, if
the potential is of such a nature that being at
the position $a$ has quite a small probability
in equilibrium? We shall show
that in these circumstances it is to be expected
on quite general grounds that this probability
has the behavior outlined above, namely a
$t^{-3/2}$ initial decay, followed by a plateau
and finally ending in an exponential decay.
At the intuitive level, this can be seen as follows:
Near $a$, the process can be viewed as
a diffusion with a drift away from $a$.
This means that the distribution of first return times
is of the type $t^{-3/2}e^{-t/t_c}$. This distribution, however,
is not normalized to one, and the missing probability
corresponds to particles that diffuse away to infinity.
These particles, however, will not really diffuse to infinity,
but rather reach the equilibrium distribution
$\exp(-V(x))$. Under these circumstances, it will from then on
reach $a$ at a rate $\exp(-V(a))$. This therefore yields
an exponential decay in the first passage probability on a
time scale $T$ much larger than $t_c$, hence yielding
the expected plateau. A more formal derivation is found in
the Appendix.

\section{Final Remarks}

\begin{figure}
\centering
\includegraphics*[width=\figwidth,angle=270]{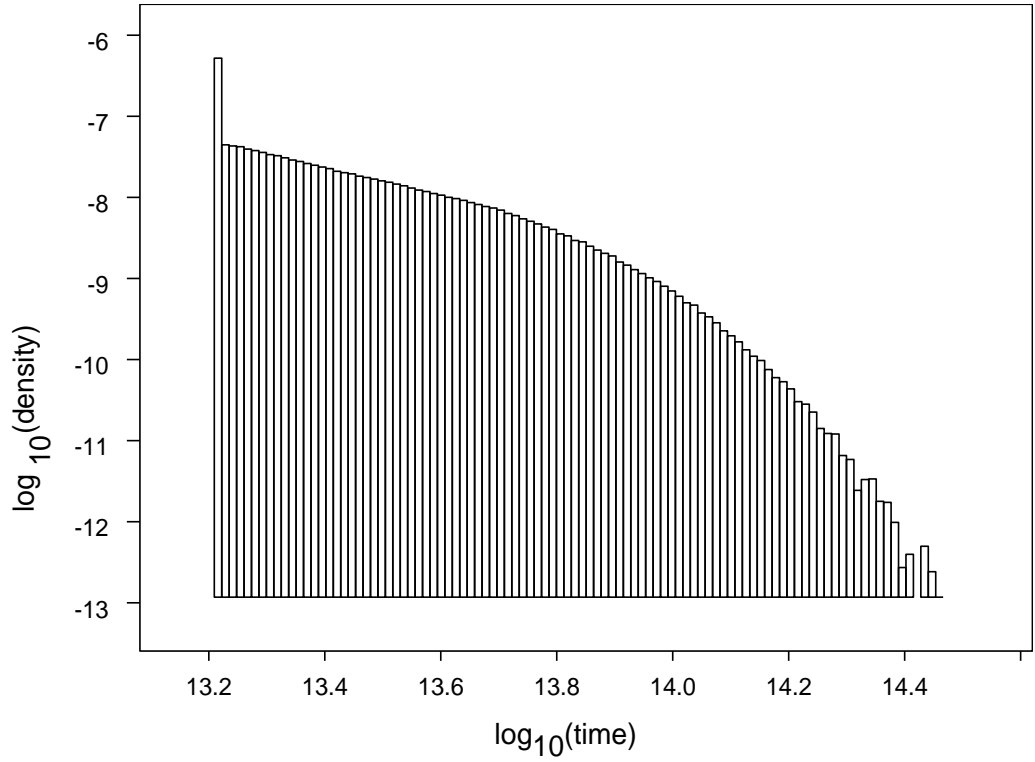}
\includegraphics*[width=\figwidth,angle=270]{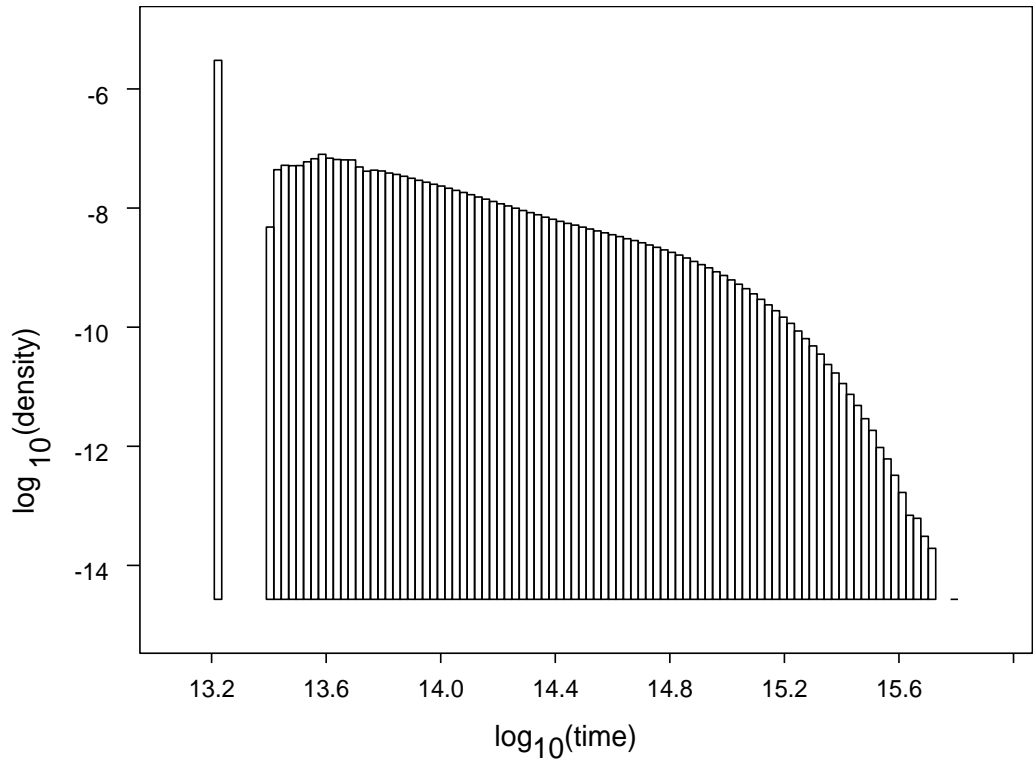}
\caption{True time delays. Log Log plot.
Top: strong coupling, bottom: weak coupling.}
\label{LogLogTime}
\end{figure}

We finally display the true time delays in the two cases of strong and
weak coupling, that is, we compute the actual time of each
Rydberg excursion.
We note in Fig.~\ref{LogLogTime}
that the different regimes detected
in the iteration plots become quite blurred. This is due to the fact
that two very similar trajectories near threshold have greatly
different excursion times as the latter diverge at this threshold.
The first consequence is that time curves span only two orders
of magnitude whereas iteration curves span eight orders of magnitude.
The first channel in the strong case and the first isolated peak
in the weak coupling case correspond to electrons which are stabilized
in the first collision and ejected exactly in the second. These
features depend very critically on the exact position with respect
of threshold of the incident ring of particles, whereas the remaining
of the picture as well as iteration curves are stable against
such minor variations. The $t^{-3/2}$ behavior in the weak
coupling case corresponds only to a longer decreasing plateau,
the slope of which is no more 1.5. The delay between $t^{-3/2}$
and exponential decay is no more visible, And finally
the last $t^{-n}$ decay has also disappeared.
If we want to obtain information about the process, we conclude
that the true time delays do not yield this information in a clear
cut fashion.

We may conclude that our model displays typical properties
of compound behavior in many respects. Yet the remarkably complicated
structure of the time delay indicates that more information can be
obtained than is commonly assumed e.g. in nuclear
compound reactions. The diffusive model seems promising, as it can
readily describe
situations where the regimes are not as clearly separated, which may
be more common.

As our analysis is limited to classical mechanics it remains an
interesting open question, how this situation translates to quantum
mechanics. Also one will have to ask what features are very model
dependent and what features may be found in more general context.
In particular we think about the possibility of a diffusive description
of pre compound situations.

\begin{appendix}
\section{Appendix}
In this Appendix, we shall show at a more
formal level that the behavior of the type
described in the text is quite general
for first return probabilities in a potential that
goes to infinity when $|x|\to\infty$.
The master equation satisfied by the
probability $P(x,t)$ of reaching $x$ at time
$t$ is given by
\begin{equation}
\frac{\partial P(x,t)}{\partial t}
=\frac{\partial}{\partial x}
\left[
\frac{\partial P}{\partial x}+
V^\prime(x) P
\right]\equiv-L P,
\label{eq:master}
\end{equation}
where the last equality defines the operator $L$.
Here the initial condition is given by
\begin{equation}
P(x,0)=\delta(x-a).
\end{equation}
From this one obtains
\begin{equation}
P(a,t)=\sum_{m=0}^\infty\frac{|\phi_m(a)|^2}{\phi_0(a)}e^{-E_mt},
\label{eq:sum}
\end{equation}
where $E_n$ and $\phi_n(x)$ are the eigenvalues
and eigenfunctions of $L$ respectively. The division
by $\phi_0(a)$ is related to the fact that $L$ is not self-adjoint
with respect to the ordinary scalar product, so that
the $\phi_n(x)$ are not orthogonal. However, replacing
$\phi_n(x)$ by $\phi_n(x)/\sqrt{\phi_0(x)}$
restores self-adjointness.

As is well-known, the Laplace transform $Q(s)$
of the first passage probability $q(t)$ is
given by\cite{RandomWalk}
\begin{equation}
Q(s)=\frac{P(s)}{1+P(s)},
\end{equation}
where $P(s)$ is the Laplace transform of
$P(a,t)$ given by (\ref{eq:sum}).

Now we need both a large and a small
time estimation for the expression
(\ref{eq:sum}). The large time
estimate is trivial: Since (\ref{eq:master})
always has an equilibrium solution, $L$ always
has a zero eigenvalue. At large times, this
eigenvalue always dominates, and one has
\begin{equation}
P(a,t)=\phi_0(a)=e^{-V(a)}.
\label{eq:sum2}
\end{equation}
For short times, on the other hand, one may simply
assume that the diffusing particle does not
go far from $a$, so that we replace $V(x)$
by its Taylor development to first order.
Then (\ref{eq:master}) can be solved exactly to yield
\newcommand\vp{V^\prime(a)}
\begin{equation}
P_d(s)=\frac{1}{\sqrt{4s+\vp^2}},
\label{eq:drift}
\end{equation}
where $P_d(s)$ is the approximation to $P(s)$
in this approximation. From (\ref{eq:sum})
follows, however, that $P(a,t)$ is
never less than $\phi_0(a)$, which we may incorporate
approximately by setting
\begin{equation}
P(s)=P_d(s)+\frac{\phi_0(a)}{s}.
\label{eq:approx}
\end{equation}
From this follows
\begin{equation}
Q(s)=\frac{\phi_0(a)\sqrt{4s+\vp^2}+s}{(s
+\phi_0(a))\sqrt{4s+\vp^2}+s}.
\label{eq:qapprox}
\end{equation}
If $s\ll\vp^2$, this reduces to a rational function in
$s$, which corresponds to a double exponential decay,
with decay times $\tau_1$ and $\tau_2$
approximately given by
\begin{equation}
\tau_1=\frac{2}{\vp\left(1+\vp\right)}\qquad\tau_2=\frac{1+\vp}
{\phi_0(a)\vp}
\end{equation}
The $\tau_2$ decay clearly dominates
the picture at those times. Since $\tau_2$ is
much larger than the time at which this approximation starts
to be valid, namely approximately $\vp^{-2}$, this implies
the existence of the plateau followed by an exponential
decay. On the other hand, in the large $s$ approximation
we may altogether neglect the correction made
in (\ref{eq:approx}) and use $P_d(s)$. This
yields
\begin{equation}
Q(s)=\frac{1}{\sqrt{4s+\vp^2}+1},
\label{eq:qlarge}
\end{equation}
which yields a $t^{-3/2}$ decay cut off by an exponential
at a time scale of order $\vp^{-2}$. This meshes therefore
quite well with the large time asymptotics, which
just then begins to set in.
\end{appendix}

\bibliography{classiqu}
\bibliographystyle{elsart-num}

\newpage

\section*{\listtablename}

\begin{table*}[hbp]
\caption{DIFFUSION PARAMETERS.
For all figures and this table: $L=50$, $J=100$,
$E=+1.275\,10^{-7}$, $2B=10^{-10}$,
initial ring radius $\alpha_0=0.0725\pi$,
$n=10^7$ iterations.
(a) Full Rydberg Case,
(b) Eliminating trapped parts of trajectories,
(c) Random $\beta$ with true Rydberg $\delta\phi+\delta\phi^\prime$,
(d) Random $\beta$ with first order $\delta\alpha$,
    no $\delta\phi^\prime$.
(e) Random $\beta$ with $\delta\alpha$ up to second order in $K$,
no $\delta\phi^\prime$.
(f) Idem including $\delta\phi^\prime$.
$t_c={\left(\frac{A_1}{A_2}\right)}^{1/\alpha_1}$
Between parentheses: one standard deviation on the last figure.}
\label{tab:fit}
\begin{tabular}{ccccccc}
\hline\hline
  &$K$&$\alpha_1$&$T$&$10^{-6}A_1$&$A_2$&$t_c$\\
\hline
(a)&0.1&1.594(2)&18735(110)&5.55(4)&70.0(6)&1184\\
   &0.2&1.779(7)& 5235(83) &6.33(7)& 397(9)& 270\\
   &1  &2.299(3)&  453(1)  &4.72(1)&13685(28)&13\\
(b)&0.1&1.604(4)&16061(225)&5.55(4)& 83(2) &1018\\
(c)&0.1&1.589(2)&21844(65) &5.50(3)&58.8(4)&1344\\
(d)&0.1&1.532(2)& 4748(70) &3.43(2)&       &    \\
(e)&0.1&1.620(2)&34320(85) &4.27(3)&12.8(1)&2566\\
(f)&0.1&1.641(1)&19759(40) &4.31(1)&39.8(1)&1169\\
\hline\hline
\end{tabular}
\end{table*}

\newpage

\section*{\listfigurename}

\end{document}